%% file: paper.tex
\documentclass[sigconf]{acmart}

\usepackage{graphicx}
\graphicspath{ {images/} }

\usepackage{amssymb}
\usepackage{algorithm}
\usepackage{algpseudocode}
\usepackage{algorithmicx}
\usepackage{amsmath}
\usepackage{multirow}

\begin{document}
\title{SAIC: Identifying Configuration Files for System Configuration Management} % TODO: replace with your title

\if0
% TODO: replace this section with code generated by the tool at https://dl.acm.org/ccs.cfm




% -- end of section to replace with generated code
\fi

\author{
Zhen Huang, David Lie
}
\affiliation{University of Toronto}
\email{z.huang@mail.utoronto.ca, lie@eecg.toronto.edu}

\maketitle

 % TODO: replace with your keywords

\input{macros}

\input{abstract}
\input{intro}

\input{problem}

\input{study}

\input{implementation}

\input{evaluation}

\input{related}
\input{conclusion}

\bibliographystyle{ACM-Reference-Format}
\bibliography{bibfile}

\input{appendix}

\end{document}

%% file: macros.tex
%\renewcommand{\dbltopfraction}{1.0}
%\renewcommand{\topfraction}{1.0}
%\renewcommand{\bottomfraction}{1.0}
%\renewcommand{\textfraction}{0.2}

%\renewcommand{\subfigtopskip}{0pt}
%\renewcommand{\subfigcapskip}{0pt}

%\newcommand{\mysubfig}[3]
%{
%\subfigure[#2]{
%\ifpdf
%\includegraphics[width=#3\linewidth]{#1}
%\else
%\includegraphics[width=#3\linewidth]{#1}
%\fi
%}
%}

\newcommand{\myfig}[4]
{
\begin{figure}[tb]
\begin{center}
\ifpdf
\includegraphics[width=#4\linewidth]{#1}
\else
\includegraphics[width=#4\linewidth]{#1}
%\rotatebox{-90}{\includegraphics[width=0.8\linewidth]{#1}}
\fi
\end{center}
\vspace{-0.15in}
\caption{#2}\label{#3}
\end{figure}
}

\newcommand{\myfigwide}[4]
{
\begin{figure*}[tb]
\begin{center}
\ifpdf
\includegraphics[width=#4\linewidth]{#1}
\else
\includegraphics[width=#4\linewidth]{#1}
%\rotatebox{-90}{\includegraphics[width=#4\linewidth]{#1}}
\fi
\end{center}
\caption{#2}\label{#3}
\end{figure*}
}

%% file: abstract.tex
%\begin{abstract}
\section*{abstract}
Systems can become misconfigured for a variety of reasons such as
operator errors or buggy patches. When a misconfiguration is
discovered, usually the first order of business is to restore
availability, often by undoing the misconfiguration. To simplify this
task, we propose the Statistical Analysis for Identifying Configuration Files (SAIC), which analyzes how the contents of a file changes over time to automatically determine which files contain configuration state.  In this way, SAIC reduces the number of files a user must manually examine during recovery and
allows versioning file systems to make more efficient use of their
versioning storage.

The two key insights that enable SAIC to identify configuration
files are that configuration state must persist across executions of
an application and that configuration state changes at a slower rate
than other types of application state.  SAIC applies these insights
through a set of filters, which eliminate non-persistent files from
consideration, and a novel {\em similarity} metric, which measures how
similar a file's versions are to each other.  Together, these two
mechanisms enable SAIC to identify all 72 configuration files out
of 2363 versioned files from 6 common applications in two user
traces, while mistaking only 33 non-configuration files as
configuration files, which allows a versioning file system to
eliminate roughly 66\% of non-configuration file versions from its
logs, thus reducing the number of file versions that a user must try
to recover from a misconfiguration.
%\end{abstract}

%% file: intro.tex
\section{Introduction}
\label{section:introduction}

Configuration management is the bane of system administration.  An
application can become misconfigured due to a multitude of causes --
human error, incorrect patches or buggy software, just to name a few.
In addition, misconfigurations are not always immediately obvious and
may only affect an application when used in a certain way.  As a
result, the cause and symptom of a misconfiguration can be
separated by weeks or even months.

When a misconfiguration is discovered, often the system
administrator's immediate priority is to restore availability to the
system.  While the ideal solution is to fix the misconfiguration,
configuration debugging is a laboriously slow process. Therefore, a
more short-term and realistic goal is to roll the broken application
back to the state just before the misconfiguration occurred. Finding
this state may also provide  clues on how to find a more permanent fix
for the misconfiguration. Many systems facilitate configuration
rollback by providing some capability for recording and rolling back
the state of the file system. For example, Windows Vista comes with
Shadow Copy disks \cite{shadowcopy} and Mac OS includes Time Machine \cite{timemachine}, both of which take
block level snapshots of an entire disk.  Such systematic solutions
are frequently combined with ad-hoc ``best practices'', such as
keeping configuration files under revision control.

However, these solutions are not quite satisfactory for recovering
from misconfigurations. Versioning an entire disk or file system
consumes a lot of disk space. As a result, such systems take
snapshots, so they may miss file versions, and even then, usually can
only maintain several weeks' worth of versions.  In addition, when a
misconfiguration occurs, the user is confronted with a legion of files
and versions within which she must find the cause of the
problem. Thus, current versioning systems will only work if the cause
of the misconfiguration is recent, and still require the user to try
rolling back many different files to find the right one.

Manually maintaining versions in revision control or manual backup
reduces the number of files versioned, saving both space and effort.
However, it requires the user to know the exact location and number of
configuration files for each application.  Unfortunately,
configuration complexity has driven more and more application
developers to provide a tool or graphical interface for configuration,
rather than allowing the user to edit configuration state directly.
This effectively abstracts and hides the storage of the configuration
settings from the user.  Unlike traditional Linux/UNIX or Windows applications that store configuration in {\tt /etc} or in the Windows registry, these
applications have being storing configuration state in a variety of files that
do not have descriptive names.  Versioning file systems, which record
versions at the file system level instead of the block level, can
selectively version files on a file system, but they still place the burden
on the user to find the configuration files~\cite{ElephantFS}.

%\begin{itemize}
%\item Give approach outline here
%\item Uses a statistical approach to compute a rate of change of a
%file over a period of time.  
%\end{itemize}

Since we cannot assume that a user knows where the configuration files
for her applications are stored, we propose the Statistical Analysis for Identifying Configuration Files (SAIC), that will automatically find an application's configuration files for the user.  SAIC treats
applications as black-boxes, requiring no a priori knowledge of the
application, and heuristically ranks a file's likelihood to contain
configuration state by observing how it is accessed and how its
content changes over time.  It performs a statistical analysis of
how a file's contents change over time to determine whether the file
stores configuration state or not.  The intuition behind this approach
is that configuration state changes very slowly, so that files that
exhibit a high amount of {\em similarity} as they are modified are
more likely to contain configuration state.  Thus, by measuring the
similarity of a file's versions, SAIC can assign a score that denotes the
likelihood that a file stores configuration state.  This information
is then used to guide a versioning system into selecting which files
to keep and which to discard when it is approaching its space
quota. Having a more accurate picture of which files to version  not
only reduces the storage requirements of the versioning system, but
also reduces the effort the user must expend to roll the application
back to a working state and unravel the cause of the misconfiguration.

\subsection{Contributions}
This research makes the following three contributions to the area of configuration management:
\begin{enumerate}
\item To the best of our knowledge, it is the first research that systematically analyzes the locations, formats, sizes, and change patterns of configuration files
\item A novel statistic-based similarity metric to identify configuration files for arbitrary applications
\item A working system that automatically identifies configuration files using the similarity metric and aids a user in recovering configuration problems
\end{enumerate}

The structure of this paper is as follows. In section~\ref{section:background}, we provide the relevant background for the reader. Section~\ref{section:problem} characterizes the problem SAIC faces with identifying configuration files. Section~\ref{section:analysis} presents our trace-based analysis of configuration files. Section~\ref{section:design} and Section~\ref{section:implementation} describe SAIC's solution for identifying configuration files and which versions hold significant configuration changes. Finally, we present evaluation results in Section~\ref{section:evaluations} and conclude in Section~\ref{section:conclusions}.

%% file: problem.tex
\section{Problem Definition}
\label{section:problem}

\subsection{What is a Configuration File?}

One of the difficulties in dealing with configuration problems is that
the term ``configuration file'' is itself a nebulous term.  To resolve
this difficulty, we begin by defining a configuration file for the
purposes of this thesis.   We first introduce the term {\em application
state}, which describes persistent state that an application maintains
throughout its lifetime.  We then categorize application state as
either {\em task-dependent}, {\em time-dependent} or {\em
configuration state}.  A task is an application-specific unit of work,
such as navigating to a web page with a browser, or editing a file
with an editor.  Thus, task-dependent state reflects the sequence of
tasks an application has performed -- an editor's list of recently
opened files or the state of a browser's web cache for example.
Time-dependent state records the passage of time or counts the number
of occurrences of certain events -- such as a timer to check for
patches or perform a cleanup.   Configuration state defines the
application's behavior over time and across multiple tasks.  With this
model consisting of three types of application state, we can finally
give our definition of a configuration file:
\begin{quote}
{\bf Configuration File:} A file that stores application state that is
neither task-dependent nor time-dependent, and whose contents may thus
affect an application's availability or performance over a period of
time and across a number of tasks.
\end{quote}
We note that misconfigurations do not necessarily affect {\em all}
tasks because not all configuration state is used for every task.
Thus, while configuration state is not task-dependent, the act of
accessing configuration state is.  For example, directing a web
browser to view only web pages on the local disk would not reveal
problems stemming from  misconfigured web proxy settings.  

We found this definition useful when trying to decide whether a
borderline case was a configuration file or not.  One such case was
for files whose contents are set by a remote party as opposed to the
user, but which have a configuration-like function for the
application.   For example, Firefox periodically downloads a blacklist
of phishing sites, which it uses to protect the user.  This file's
contents are neither task- nor time-dependent, but its contents can
affect the usability of Firefox by  preventing the user from visiting
even legitimate websites if incorrectly configured.  As a result, our
definition considers such files configuration files.  If this
blacklist does becomes misconfigured, rolling this blacklist back to
an earlier version will fix the problem -- though possibly only
temporarily.  In addition, knowing that this file is the cause of a
misconfiguration is valuable to the user.  Thus, while our definition
is broad, we feel it does contain most of the files that are of
interest to system administrators who are trying to debug a
misconfigured application.  We concede that this definition also does
not cover every possible source of misconfiguration.  For instance,
applications may derive their configuration dynamically from a device
or network resource without actually writing it to the file system.
Since SAIC can only observe configurations saved to disk, it will
not be able to analyze these cases.  We leave the debugging of such
configuration problems outside the scope of this thesis.

In addition, an application does not always store all of its configuration state in configuration files. Many applications provide default values for their configuration parameters. If a configuration parameter has a default value, the applications often do not store the value in their configuration files, since the default value of a configuration parameter can be defined as a constant in the application's code and thus would probably be saved into the application's executable file. It is also possible that the default value of a configuration parameter is implied by the logic of the application's code. Thus the configuration state of an application includes not only the configuration states that are stored in the configuration file, but also those stored directly as constants or indirectly as code in executable files. 

\subsection{Identifying Configuration Files}

One may ask why it is difficult in the first place for a user to
identify the configuration files of an application.  For applications
whose configuration files are not meant to be edited manually, the
application developer has no reason to make the location of the
configuration files obvious to the user. As a result, relying
exclusively on file names is unreliable.  For example, Firefox stores
configurations in files such as {\em pluginreg.dat, secmod.db,
prefs.js, localstore.rdf} and {\em mimeTypes.rdf}.  The Macromedia
Flash plugin maintains several files named {\em settings.sol} in
various directories, but only one of them actually contains
application configuration data (the others contain website-specific
data).  In addition, it appears that for portability purposes, both of
these applications do not use OS-specific conventions for the storage
of their configuration state. For example, rather than using the
registry in Windows, Firefox stores all its configuration state in a
subdirectory under the user's home directory, just as it does on Linux.

Another problem is that configuration files often have a sprinkling of
time-dependent and task-dependent state mixed in with the
configuration state.  For example, Firefox stores timestamps of
events, the period between updates, and the exit status of the
previous execution (i.e. exited cleanly or crashed) in the {\em
prefs.js} configuration file. Similarly, the GNOME desktop system
records timestamps in each GNOME application's {\em \%gconf.xml}
configuration files. As a result, it is more appropriate to score
files by the proportion of configuration state they contain, rather
than try and classify files as entirely configuration or
non-configuration.  In addition, while the vast majority of the
content in such files is configuration state, the vast majority of
modifications to the files are actually to these sprinklings of
non-configuration state.  This adds a great deal of noise to the file
system interaction between an application and its configuration files.

Finally, we note that configuration files are not always accessed in a
predictable pattern.  While many configuration files are read at
start up~\cite{mirage}, some applications read their configurations
on demand.  For example, Firefox will only access the {\em
pluginreg.dat} configuration file if plugins are used or modified.
Accesses to such configuration files occur neither in every execution
of an application, nor do they occur at the beginning of the
execution.  As a result, any heuristic that assumes configuration
files are always read at start up will miss configuration files.

\subsection{Misconfiguration Recovery}
\label{sec:problem:recovery}

Unfortunately, without knowledge of which files contain configuration
state, recovery from a misconfiguration is a very expensive task.  To
illustrate, suppose an application has failed due to a
misconfiguration and the user would like to determine which files read
during the failing run are causing the problem.  The user may profile
the failing application to determine which files it accesses, and then
query the versioning system to find all versions of those files.  She
may then start rolling back various files, one version at a time, to see
if the problem is fixed.  We note that while one could use binary
search to speed up the task of finding a working configuration
state~\cite{chronus}, binary search can only identify a point where an
application transitioned from a working state to a broken state.
Unfortunately, this has the disadvantage that if there was more than
one time in which the application's configuration state was not valid,
then it cannot guarantee that the working file version it finds is the
most recently working state.  Since configuration files are not
modified atomically, there can be many transient periods when the
configuration files are in an invalid state.  Thus, the only way to
ensure that the most recent working configuration state is found is
to do a linear search backwards in time from the current state of the
application.

\begin{table*}[t!]
\center
\begin{tabular}{|l|r|l|r|r|r|r|} \hline
\multicolumn{1}{|c|}{\bf Trace} &
\multicolumn{1}{|c|}{\bf Days} &
\multicolumn{1}{|c|}{\bf Application} &
\multicolumn{1}{|c|}{\bf Non-code files}  &
\multicolumn{1}{|c|}{\bf Versioned files} &
\multicolumn{1}{|c|}{\bf Config}  &
\multicolumn{1}{|c|}{\bf Versions (config/total)} 
\\ \hline \hline
%\multirow{2}{*}{1} & \multirow{2}{*}{22} & Firefox & 1160 & 1010 & 30 & 6260/224154 \\
\multirow{4}{*}{1} & \multirow{4}{*}{22} & Firefox & 851 & 507 & 7 &
6485/295841 \\
& & GNOME & 1552 & 392 & 16 & 321/2064 \\ 
& & Flash & 37 & 17 & 1 & 7/130 \\ 
& & VMware & 337 & 189 & 6 & 160/2926 \\ \hline
%& & GNOME & 972 & 1224 & 11 & 296/1207 \\ \hline
%& GNOME & 972 & 1224 & 11/25 & 296/1207 \\ \hline
%\multirow{2}{*}{2} & \multirow{2}{*}{12} & Firefox & & & & x/x\\
\multirow{4}{*}{2} & \multirow{4}{*}{12} & Firefox & 1247 & 568 & 15 & 234835/751808 \\
%& GNOME & 1005 & 1160 & 8/128 & 73/1993 \\ \hline
%& & GNOME & 1005 & 1160 & 8 & 73/1993 \\ \hline
& & GNOME & 2537 & 493 & 21 & 13298/30388 \\
& & Flash &  53 & 35 & 1 & 15/1995 \\ 
& & JEdit & 6876 & 73 & 2 & 46/5002 \\
& & Amarok & 3086 & 89 & 3 & 43/4782 \\ \hline
\end{tabular}
\caption{Application file usage measurements.  For each trace, we give the
number of non-code files accessed by the application, the number of
files that have been modified and have versions stored on the system,
and the number of configuration files the application uses. The last
column gives the aggregate number of configuration file and total file
versions created by the application over the entire trace.
}
\label{tbl:app_trace}
\end{table*}

To quantify the difficulty of this task, we collected system traces
from two Linux workstations in our lab.  The traces were collected
with a kernel module that intercepts system calls, versions files
using a redo log, and records which applications access each file.  A
new file version is created on every write system call or memory map
of a file.  In cases where several contiguous writes are made to a
file, this is counted as a single version.  Table~\ref{tbl:app_trace}
gives data extracted from the traces on several applications that were
used by the workstation users.  GNOME represents the GNOME suite of
desktop applications, Flash is the Macromedia Flash plugin, VMware is
VMware workstation, JEdit is a JAVA-based editor and Amarok is an
open-source music player.  We describe the applications in more detail
in Appendix~\ref{sec:appendix:apps}.  We note that Trace 1 has less
activity than Trace 2, despite a longer trace period, simply because the
user did not use the machine as much.  To determine whether a file
that was accessed contains configuration state, we use a combination
of application documentation and profiling of the application.

The traces show that without knowledge of which files may contain
configuration state, the user would have to sift through hundreds of
files that were accessed and potentially try thousands of versions to
find the most recently working configuration state.  Instead of having
to examine each versioned file, knowing which files are configuration
files reduces the number of files the user has to consider by one to
two orders of magnitude.  The number of file versions to try  is also
reduced by an order of magnitude or more in most cases and by at least
a factor of 2 in the worst case.

%,  except in the case
%of Firefox in Trace 1 where it was only reduced by 56\%.  This can be
%explained by the fact that the user in Trace 1 was developing a web
%application during the course of the trace and thus used Firefox as a
%development and debugging tool, which resulted in more configuration
%changes.

%Ocasta simplifies the configuration problem in two ways.  First,
%Ocasta indicates which files and file versions may contain
%configuration information, reducing the effort the user must expend
%searching for the most recently working configuration state.  Second,
%by providing such information to a versioning file system, Ocasta
%allows a longer history of versions to be stored, allowing the user to
%recover from misconfigurations that occured a longer time in the past.
%To characterize this problem, we will measure the number of files
%applications read on a typical run, which gives an indication of the
%effort a user must expend searching for which file to rollback.  We
%then measure measure how much space it would take to version every
%file that is modified on a system over a period of time to give some
%indication of how much space Ocasta can save a file system that is
%trying to create a long term archive of configuration states.

%To measure the number of files applications access, we record every
%file that is read by Firefox and on a typical execution and present the results in
%Table~\ref{tbl:app_trace}.  We then manually count the number of
%configuration files that were accessed during the execution. {\bf More
%to come when we have the results...}

\begin{table}[t!]
\center
\begin{tabular}{|p{0.1in}|r|r|p{0.6in}|r|r|} \hline
\multicolumn{1}{|c|}{\bf Trace} &
\multicolumn{1}{|c|}{\bf Days} &
\multicolumn{1}{|c|}{\bf Write ops} &
\multicolumn{1}{|c|}{\bf Bytes} &
\multicolumn{1}{|c|}{\bf Files} \\ \hline \hline
1 & 22 & 7,884,155 & 3.01GB & 12,983 \\ \hline 
2 & 12 & 62,647,496 & 21.28GB & 96,470 \\ \hline
\end{tabular}
\caption{Versioning storage overhead measurement.  We give the number
of write operations, bytes written and number of files modified in
each trace.}
\label{tbl:version_trace}
\end{table}

In addition, versioning only files that are likely to contain configuration state can reduce storage overhead.  
We estimate the amount of storage necessary to version each of the traces using the size of the
redo logs from the trace.  As shown in Table~\ref{tbl:version_trace},
both traces would require a significant amount of storage to version
all the data written to the file system over the length of the trace.
While one may argue that storage is cheap, the user effort required to
manage the storage and analyze large amounts of data in the storage is
expensive. Therefore, the ability to separate configuration
state from other types of application state will be beneficial to
system configuration management.

%% file: study.tex
\section{Configuration File Analysis}\label{section:analysis}
To the best of our knowledge, no previous work has systematically analyzed configuration files. However, we believe this kind of analysis is necessary to discover an effective and efficient method in identifying configuration files. In light of that, we analyzed various aspects such as format and change patterns of configuration files for 7  popular Linux applications. The goal of our analysis is to find out how configuration files differ from non-configuration files. Our analysis aims to answer the following questions.
\begin{itemize}
\item Where does an application store its configuration files and non-configuration files?
\item What are the format and size of configuration files and non-configuration files?
%\item How regular and configuration operations affect configuration files?
\item What are the change patterns of configuration files and non-configuration files?
\end{itemize}

\subsection{Methodology}
We analyze application behaviors that are related with configuration operations in a way similar to black-box testing. We perform experiments with applications by running them in a controlled environment to collect the traces of file accesses and file system changes made by the applications. Then we study the file system traces from different perspectives in order to understand the nature of configuration files.

\subsection{Trace Collection}
We developed a versioning file system from the code base of Forensix \cite{Goel2008} to collect the trace of each application. The versioning file system intercepts file system calls and stores every file system change into versioning logs. Each version log contains versions for one unique file. The versioning file system creates a new version for every write operation to a file. It also keeps the history of the file system calls made by the application. The file system call history is used to determine which files are accessed or updated by which applications, and when these accesses or updates occur.

To compare how the files are changed with and without configuration related operations, we performed both configuration operations and other regular operations with the applications. To be able to easily identify which file changes are related to configuration operations and which file changes are not, we divided our operations for each application into two phases: a non-configuration phase and a configuration phase. We first perform regular operations such as browsing web pages in a web browser or editing documents in a text editor in the non-configuration phase. We do not perform any configuration related operations during this phase. To make the application generates every possible file change, we execute a variety of different operations with the application. Furthermore, we execute the operations long enough to ensure the application produce a reasonably long history of file changes. 

We perform configuration operations in the configuration phase after we finish the non-configuration phase. In the configuration phase, we try to make every possible configuration change by modifying the default value of every configuration parameter of the application from the interface provided by the application. These changes should be reflected in file system changes that are isolated from the file system changes in the non-configuration phase by the time of the changes. %We ensure our tasks generate a reasonable number of versions for the files updated by the application. 

\subsection{Trace Analysis}\label{sec:trace-analysis}

%\subsubsection{File Change Patterns}
An important objective of our trace analysis is to compare the change patterns of configuration files and non-configuration files. We achieve this by tracking the change history of each individual \textit{data entry} in a file. A data entry can represent either a configuration parameter or a non-configuration parameter. We designed a simple split-and-match algorithm to approximately identify an individual data entry in a file. 

The algorithm first splits a file version into unique \textit{data fragments} by certain specific delimiters. A data fragment is a series of contiguous bytes in the file version. It is a particular instance of a data entry. For example, one configuration file (prefs.js) of Firefox can have two data fragments, each in a different version. One data fragment is ``user\_pref(``javascript.enabled'', true);'' and the other data fragment is ``user\_pref(``javascript.enabled'', true);''. These two different data fragments refer to two different values (true and false) of the same data entry that represents the configuration parameter with name ``javascript.enabled''.

For this study, we choose to use line feed as the delimiter to split a file version and analyze data changes of ASCII files only. As we will shown later, almost all configuration files are ASCII and they usually are organized into lines of text, which contains the name and the value of a configuration parameter. Many non-configuration files are also ASCII. So line feed can be used as a delimiter for most configuration files and many non-configuration files. 

To track the change history of a data entry, we need to find the data entry to which a data fragment refers to. This is necessary since a data entry change corresponds to two different values of the data entry and thus two different data fragments. Failing to match the two different data fragments that refer to the same data entry, the data entry change will be recognized as two unrelated events: a removal of a data entry and a insertion of another data entry. Thus the data entry change will not be identified correctly.

To solve this problem, we use longest common substrings algorithm \cite{clr} to match two different data fragments that refer to the same data entry. One issue of this method is that two different data fragments that refer to two different data entries may be incorrectly matched if they happen to have a long common substring. We use two orthogonal approaches to eliminate this kind of incorrect matches. First, a match can only be made between a data fragment that is \textit{removed} from a file version, and a data fragment that is \textit{added} to the same version. This is because a data entry change means the data fragment refers to the old value of the data entry will be replaced by the data fragment refers to the new value of the data entry in the file version. In addition, two added data fragments or two removed data fragments can not refer to the same data entry, because this means the version contain two different values of a same data entry, which should not happen for configuration files. Second, we observed that a configuration file can not have two data entries that represent the same configuration information. Otherwise, the meaning of the configuration information will be confusing. Therefore, two different data fragments can not be matched if they ever co-exist in a same file version.

Based on our split-and-match algorithm, we developed a utility to analyze versioning logs that are generated by our versioning system. This utility works on one versioning log at a time. It extracts every version of a file, splits each of them into data fragments, match a data fragment to a data entry, and builds a database of these data entries. It tracks every change to each data entry. The utility collects information on each data entry including how many times the data entry is added in a version, is deleted from a version, is changed from one version to the next version, how many versions the data entry exists in, and the first time the data entry is added, deleted, or changed. This kinds of information describes the change patterns of the data entries in a file. 

%\subsubsection{Configuration Files and Non-configuration Files}
From the system call history, we find out which directories and files are accessed and updated by which application. For each application, we check which files it accessed contain data entries that are changed only during a configuration phase. These files are candidate configuration files since we make only configuration changes in a configuration phase. To determine which files are exactly configuration files, we manually rollback these candidate files one by one to the beginning of the configuration phase, and check if the rollback undo the configuration changes we have made in the configuration phase. If the rollback of a file undo one or more configuration changes, the file is identified as a configuration file. If the rollback of a file does not undo any configuration changes, this file is identified as a non-configuration file. All the files accessed by an application which do not contain data entries that are changed only during a configuration phase will be identified as non-configuration files.

\subsection{Analyzed Applications}
We choose 7 popular Linux applications, as shown in Table \ref{tbl:file-format}. They consist of 2 Web browsers, 3 document editors, 1 media player, and 1 Internet messenger. We intentionally select only desktop GUI applications since it is difficult to identify their configuration files even manually. The software vendors of these application usually provide little documentation on which files are configuration files.

\subsection{Analysis Results}
Our experiments with these applications are performed on a VMWare virtual machine that runs Ubuntu 8.10 Linux with kernel 2.6.27. We choose to use a virtual machine monitor since it is easier for us to rollback the system to a previous snapshot to undo any unwanted changes made to the system. In this section, we present our analysis on the location, format, size, content, and change patterns of the configuration files of our chosen applications. It should be noted that the configuration files and non-configuration files that we discuss in this section refer to those files of the applications under our analysis. Specially, the non-configuration files refer to only the non-configuration files that are stored in the same locations of the configuration files. We do not consider non-configuration files which are accessed by the applications and are stored in other directories for this study.

\begin{table}
\centering
%\resizebox{1\columnwidth}{!}{
\scalebox{0.9}{
\begin{tabular}{|c|p{0.8in}|p{0.7in}|p{0.7in}|}
\hline  {\bf App.} & {\bf Config Files Location} & {\bf Config Files (ASC/BIN)}  & {\bf Non-config Files (ASC/BIN)}
\\ \hline
\hline  Firefox & .mozilla & 10 & 3/16\\ 
\hline  jEdit & .jedit & 2 &  9  \\ 
\hline  Amarok & .kde & 1 &  0  \\ 
\hline  Opera & .opera & 3/1 &  8/13   \\ 
\hline  Texmaker & .config & 2  & 0    \\ 
\hline  Openoffice Word & .openoffice.org2 & 12 &  0  \\ 
\hline  Skype & .Skype & 1 & 9  \\ 
\hline 
\end{tabular} 
}
\caption{Location and format of configuration files. The locations of configuration files are all relative to a user's home directory. It shows the number of ASCII and binary configuration files of each application. It also presents the number of ASCII and binary non-configuration files. }
\label{tbl:file-format}
\end{table}

\subsection{File Location and Format}
As shown in Table \ref{tbl:file-format}, there are 32 configuration files for the 7 applications in total. All applications store their configuration files under some hidden directories in the user's home directory. Three applications have more than one configuration files. All except one of the configuration files are in ASCII format including XML format. %In these files, each line in the file usually represents one data entity, which can be a configuration parameter, non-configuration data, or a line of comment. 

Only one application does not have any non-configuration files under the same directory that contains its configuration files. There are 58 non-configuration files under the same directories of configuration files. Half of the non-configuration files are also in ASCII format.
%?? of them uses some format of tags to wrap the data, which is similar to the XML format. ?? files sometimes uses more than one line to represent a configuration parameter. (i.e. Adobe reader)

\myfig{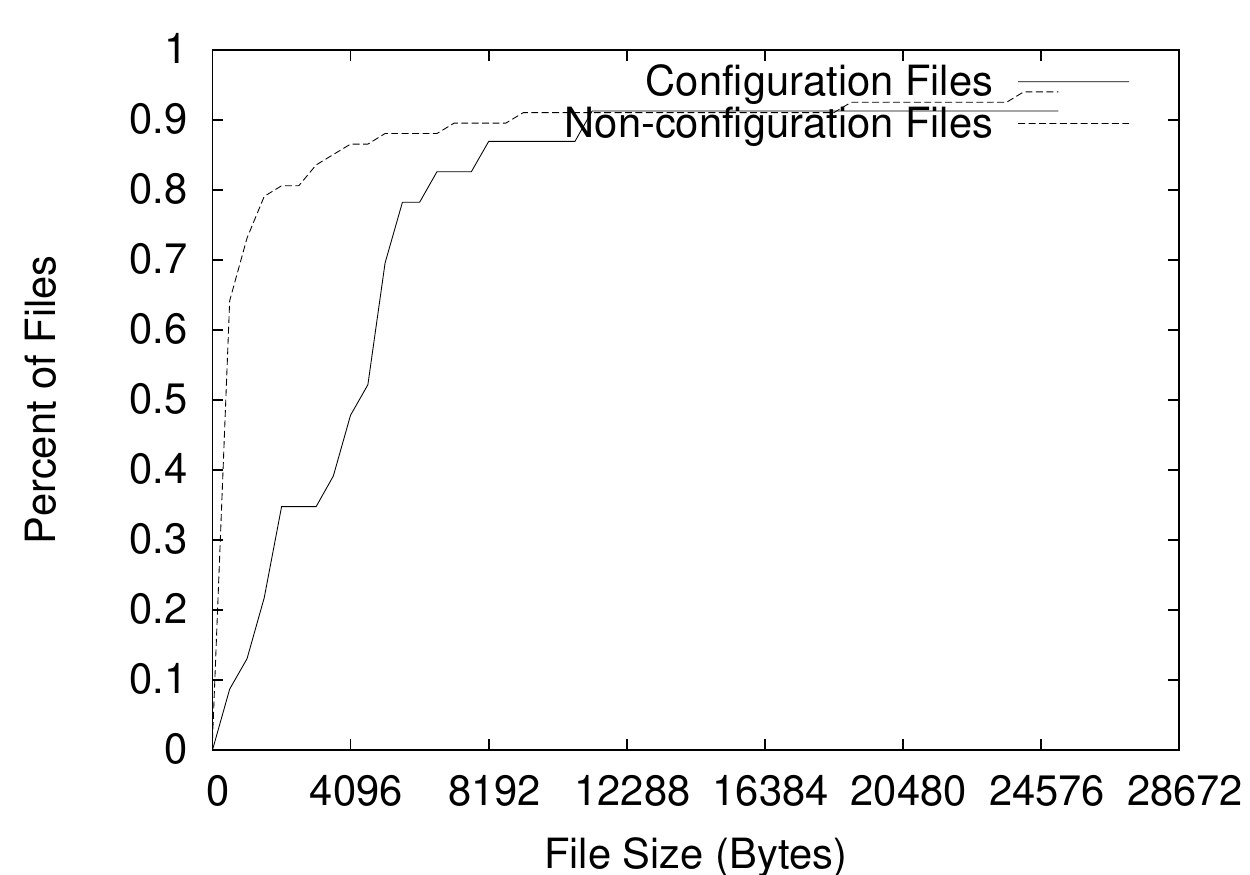}{Comparison of the sizes of configuration files and non-configuration files.}{fig:file-sizes}{0.85}

\subsection{File Size}
Figure~\ref{fig:file-sizes} shows the sizes of configuration files and non-configuration files for all the applications. The sizes of the configuration files are distributed in three ranges of sizes: below 2KB, between 2KB and 5KB, and above 5KB. Around 35\% of configuration files are less than 2KB, while 22\% of the configuration files are larger than 5KB. In contrast, most of the non-configuration files are very small. About 80\% of non-configuration files are smaller than 2KB, although the largest files are non-configuration files. %need more stuff to discuss, like type of small files, etc.

\subsection{Data Entry Type}
We classify the type of data entries in a configuration file into either configuration state or non-configuration state. This data entry type has no relation with the concept of programming language data type such as binary, number, or character string. As described in Section \ref{sec:trace-analysis}, we identify configuration files by identifying data entries changed only during a configuration phase. These data entries are called \textit{configuration data entries} and they make up the configuration state of a file. The percent of configuration state in a file is defined as the ratio of the number of configuration data entries to the total number of data entries of a file. %Since the complete configuration state is the collection of the values of an application's configuration parameters, one configuration state data represents one configuration parameter.

\myfig{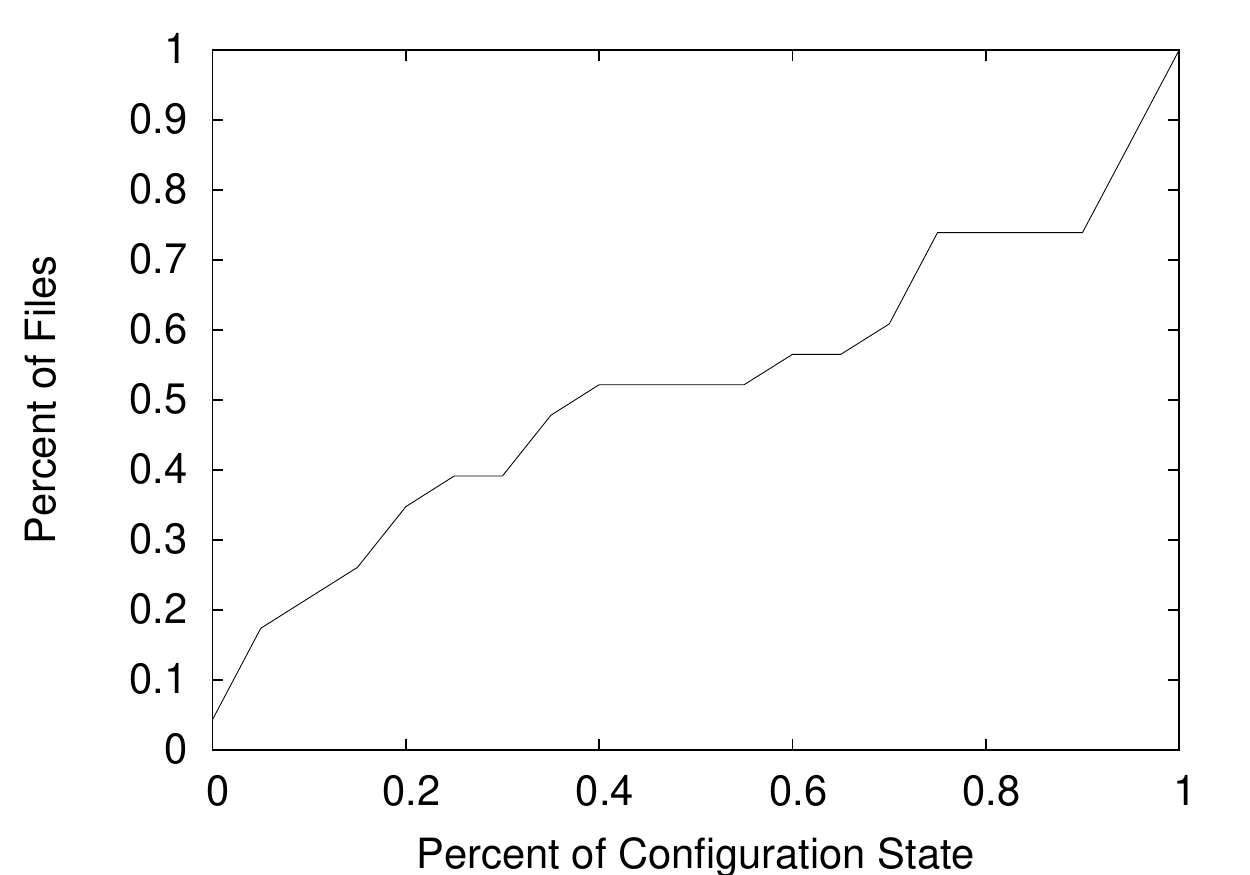}{Percent of configuration state in all the configuration files.}{fig:config-ratios}{0.85}

Figure~\ref{fig:config-ratios} shows the percent of configuration state in each configuration file. Half of the configuration files (52\%) contain less than 40\% configuration state, and 25\% the configuration files contain more than 72\% configuration state. In addition, 30\% the configuration files are almost entirely composed of configuration state. This indicates that configuration files contain data related to both configuration state and non-configuration state. Looking into the configuration files, we found that the non-configuration state includes timestamps, list of recently opened files, last user actions, window size of last execution, temporary file name, debugging information, and so on.
 %One reason that many configuration files contain less than 40\% configuration state is probably that we missed changing some configuration parameters during our experiments. Thus the data entries represent information on these configuration parameters are not identified as configuration data entries. This is not surprising as many of these applications provide a large number of configuration parameters. Some configuration parameters can only be changed when some other configuration parameter has a certain value.

\myfig{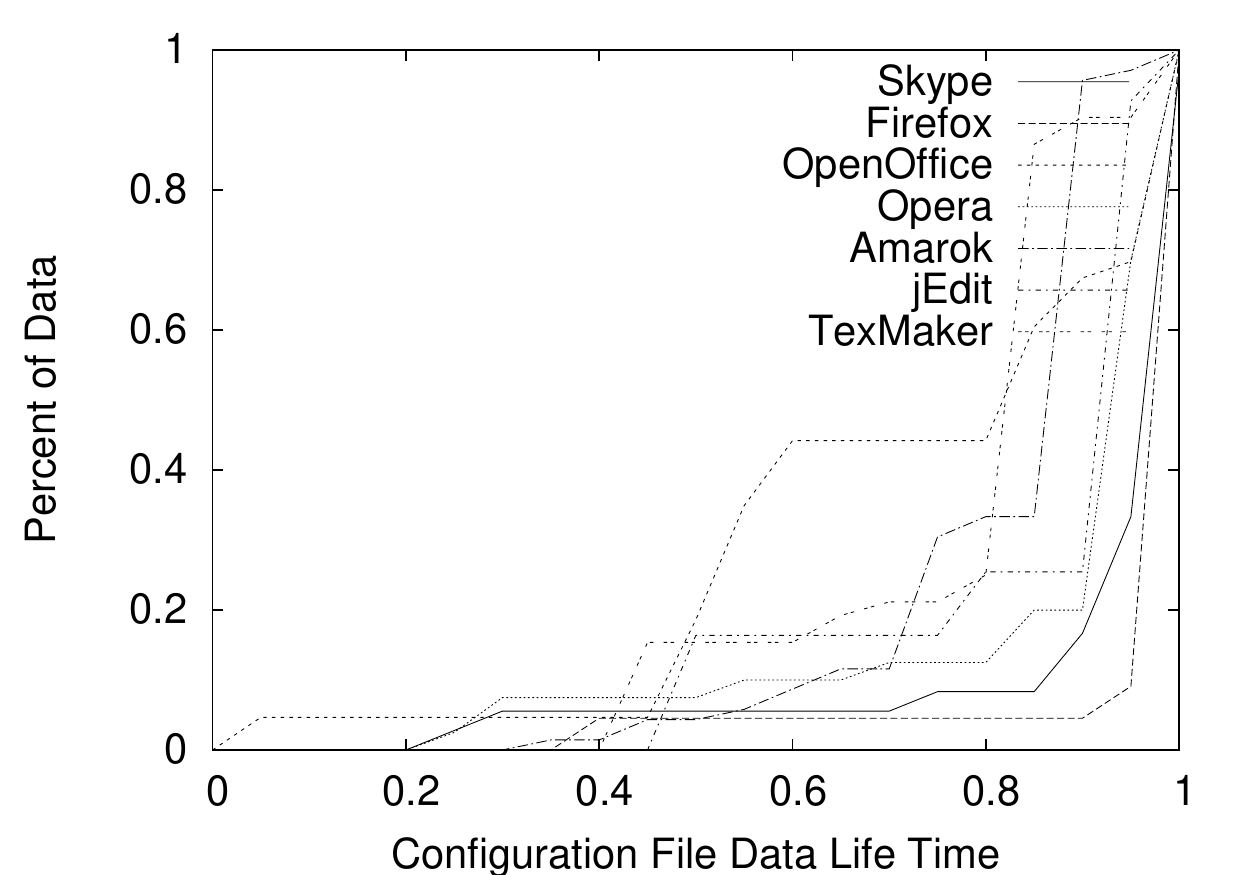}{Life time of data fragments in configuration files.}{fig:config-lifetimes}{0.85}
\myfig{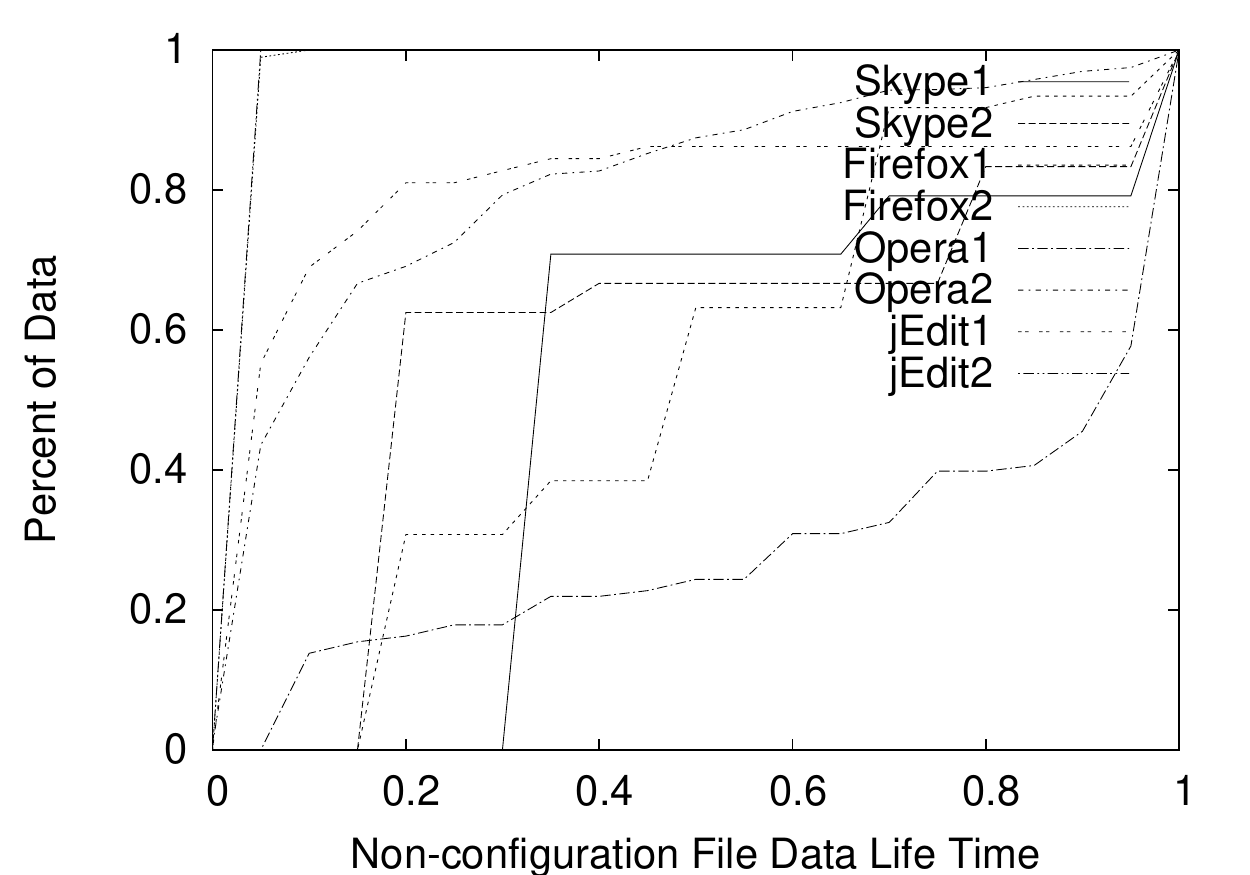}{Life time of data fragments in non-configuration files.}{fig:nonconfig-lifetimes}{0.85}

\subsection{Data Fragment Life Time}
We measure how much data in a file is changed across the different versions of a file by using the life time of the data fragments of the file. A data fragment's life time is the ratio of the number of versions in which the data fragment appears, to the number of the total versions of the file that contains the data fragment. We choose the configuration file with most versions for each application. Because some applications do not have non-configuration files under the directories in which their configuration files exist, we choose two non-configuration files for each of the applications that have non-configuration files. Figure~\ref{fig:config-lifetimes} and Figure~\ref{fig:nonconfig-lifetimes} present the data fragment life time of the configuration files and the non-configuration files, respectively. For the configuration files, all except one of the curves start low and remain flat from the short life time side, and rise steeply near the longest life time. This means most data fragments in the configuration files have a long life time.  On the contrary, the curves of the non-configuration files generally rise steeply at the short life time side, which means most data fragments in the non-configuration files have a short life time. 

%\subsection{Data Change Ratio}
%We also studied the change ratio of each chunk in a file. Change ratio of a chunk is defined as .... Figure ?? presents the change ratio of each chunk in a file. The chunks that represent configuration state have a change ratio of ??. The other chunks have a change ratio of ??. The change ratios of these two kinds of chunks have a big gap.

\subsection{Discussion}
The analysis of configuration files illustrates both challenges and promises of our research in identifying configuration files. First, applications tend to store many non-configuration files in the same location where their configuration files are stored, although configuration files are all stored in some hidden directories in the user's home directory on Linux. In addition, both configuration files and non-configuration files can be either ASCII or binary. This makes it difficult to identify configuration files by their locations or formats. Second, configuration files usually contain both configuration related data and non-configuration related data. It is thus more appropriate to score configuration files by the proportion of configuration related data they contain. Finally, the life time of data fragments in a configuration file is much longer than those in a non-configuration file. This indicates that configuration files change much more slowly than non-configuration files. This difference between configuration files and non-configuration files probably can be leveraged to identify configuration files.

%\begin{itemize}
%\item configuration files are ASCII. We can use string match to find option name in a file chunk, and thus be able to match changed chunks.
%\item configuration files contain both data represents configuration state and data represents other application state.
%\item similarity can be used to identify configuration files since a) most chunks in configuration file have relatively long life time. b) configuration states have low change ratio.
%\item non-configuration files usually stores in the same directory, can't use directory to identify file type; there are more nonconfig files (57) than config files (23)
%\item Applications use write as whole method to update configuration files, so it is hard to find which chunk is changed by logging system calls.
%\item Sometimes applications will change the order of the chunks in a configuration file. Thus a diff between two adjacent versions of a configuration file can not tell exactly which chunk is changed.
%\item Some applications have default or reset button in the user interface to change config options, however it is not desirable for the user in many cases. For example, there is no default value for proxy server.
%\item Many applications will continue to work when their configuration files are removed. In this case, they usually regenerate the default configuration files. But it has the same disadvantages as above.
%\end{itemize}

%% file: implementation.tex
\section{The Design of SAIC}
\label{section:design}

\subsection{System Model}
\label{sec:model}

% talk about performance, running periodically in the implementation section

SAIC has two modes of operation.  We assume that the system already
has a versioning facility to record file versions in place.  During
regular system operation, SAIC observes file system activity and
assigns a similarity score to each file that the file system versions.
When the versioning file system nears its space quota and needs to
discard file versions to make additional room, it will start by
discarding versions of files with the lowest similarity score first.  In addition,
in this mode, the resource requirements of SAIC must be kept to a
minimum since SAIC is sharing compute resources with other tasks on
the machine.  

When the user discovers a misconfiguration and wishes to initiate
recovery, she switches SAIC into recovery mode.  In recovery mode,
the similarity scores are used by the user to identify files that
contain configuration state.  The user will
begin by selecting the highest scoring files to try rolling back
first, since these are the files that contain the largest amount of
configuration state.  SAIC currently makes the simplifying
assumption that configuration files can be independently rolled back
and restricts its applicability to misconfigurations that can be
solved by rolling back a single configuration file.  A more complete
model would also try rolling back combinations of configuration files
simultaneously in order to find the most recent working state.  In
recovery mode, we assume SAIC will have more resources available to
it since this mode is run only when the user needs SAIC to help
recover from a misconfiguration.

%SAIC does not assume any application-specific knowledge, and relies
%exclusively on the file system behavior of an application to determine
%its configuration files.  To do this, 

%In addition, SAIC can also help identify
%significant  versions of a file, which are the result of configuration
%state modification.  Since these versions contain configuration state
%changes, rolling back to the immediate previous version is more likely
%to enable the user to recover from the misconfiguration.  

SAIC assigns similarity scores in two steps.  First, SAIC applies
a set of filters to eliminate files that do not hold application
state.  These filters do not examine the contents of the files and
thus, are very efficient.  Then, SAIC computes a similarity metric
on the remaining application state files to rank them by the amount of
configuration state they contain, thus separating configuration files
from files that contain mostly task-dependent or time-dependent state.

\subsection{Filters}
SAIC uses three filters to screen out files that do not hold
application state.  By our definition, application state must persist
for the lifetime of the application.  The first filter removes files
that have been deleted by the application, which, as a result, cannot
hold state that persists for the application's lifetime.  The second
filter requires the file to be read before it is written by an
application, implying that the state held in that file persists across
executions of the application.  Any file that does not fall in this
category only contains temporary state for a particular execution
since there was no information flow from previous executions.
Examples of files removed by these filters include lock files and
files used as unidirectional communication channels between two
processes.  There is one exception to this case -- many applications
will write a default configuration file the first time they are run if
the old configuration file is deleted or if there has been a major
update.  Thus, this filter requires a file to be written before it is
read more than 20\% of the time it is opened before it is removed from
consideration.  These two filters eliminate files that only hold
temporary execution state and not application state.  The final filter
removes user data files from consideration.  For the applications we
analyzed in this paper, a simple filter that removes files that are in
user home directories and do not have a directory starting with ``.''   in
their path is able to remove all user data files.  In practice though,
a more sophisticated, domain specific filter would likely be necessary.

\subsection{Similarity Metric}
\myfig{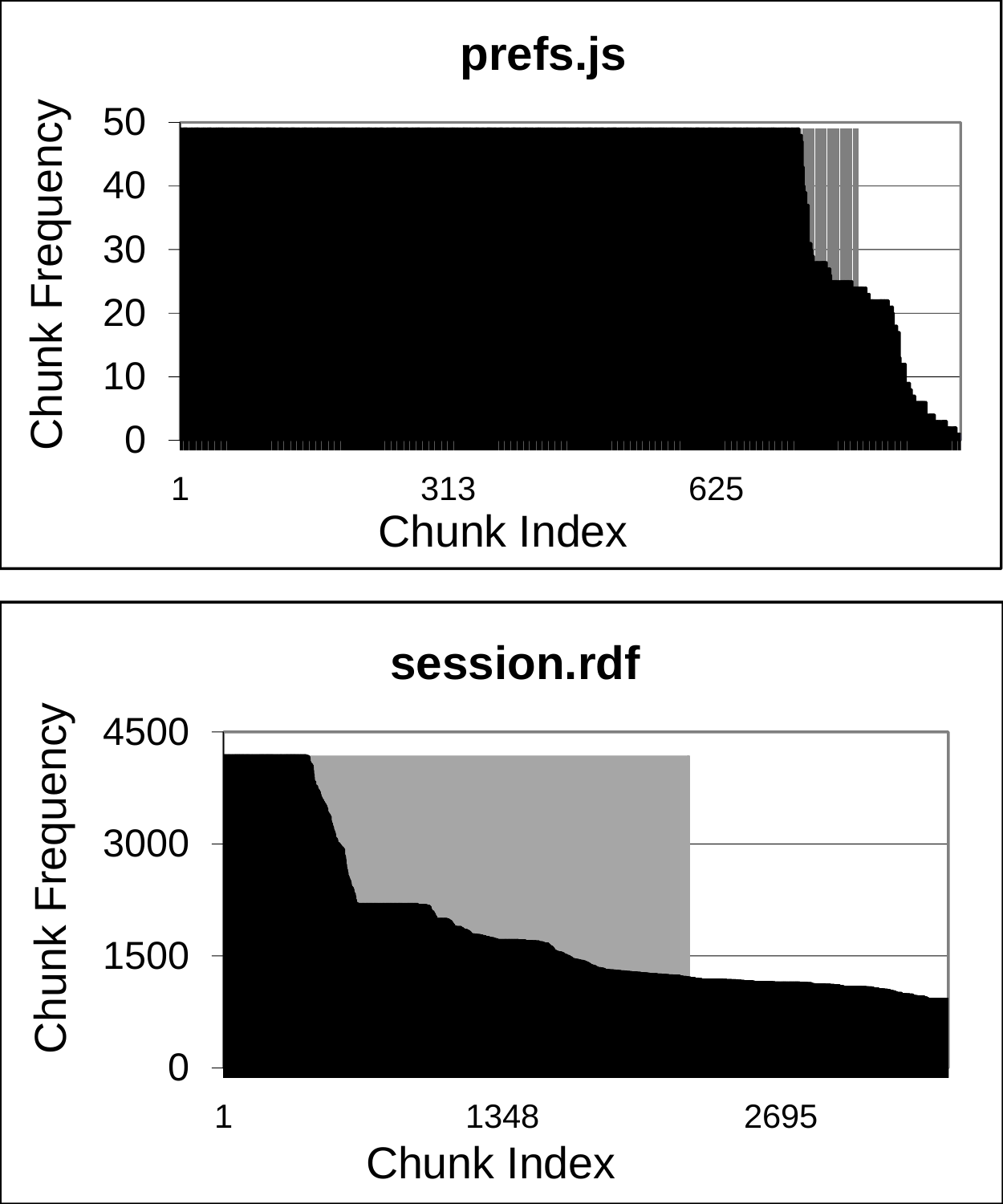}{Similarity for {\em prefs.js} (96\%) and {\em
session.rdf} (55\%).  The shaded region denotes $v \times n$ in
Equation~\ref{eqn:sim}.  Note that the lower graph has been truncated,
and actually has 17287 unique chunk values.}{fig:sim}{0.6}

After the filters have been applied, SAIC separates configuration
state from task-dependent and time-dependent state by computing the
similarity among versions of a file.  The key insight behind this
heuristic is that configuration state changes much more slowly than
both task-dependent and time-dependent state.  This is because
changing tasks and timed events occur regularly in an application's
use, but changing configuration is a rare event.  Thus, to separate
configuration state from other application state, SAIC identifies
file chunks that remain unchanged across versions of a file.  A file
that contains a high proportion of constant chunks is likely to
contain a large amount of configuration state, and thus will be
identified as a configuration file.

%Users are
%likely to change tasks frequently, and time-dependent state will
%always change regardless of what the user does.  However, users change
%configuration state rarely, much less often than they change tasks for
%example.  

To correctly capture this similarity, our algorithm must be tolerant
to insertions and deletions, since fields in configuration files may
change in size.  In addition, we have observed that applications tend
to read and parse an entire configuration file, modify it while in
memory and then periodically write out the entire configuration file.
Because of this, independent portions of the file may also be
arbitrarily reordered.  For example, the Firefox configuration file
{\em pluginreg.dat} contains information about plugins in a list
format.  The order that the plugins appear in the list has no meaning
and changes arbitrarily among versions.

To meet these requirements, we use an algorithm similar to the one
used by LBFS to find similar chunks of data for data compression in a
distributed file system~\cite{lbfs}.  Files are divided into variable
sized chunks by selecting certain offsets within the file to be {\em
anchors}.  The location of the anchors is determined by computing
Rabin fingerprints~\cite{rabin_fingerprint} on an 8 byte sliding
window over the entire file.  We then randomly select 1/16 of the
values to be anchors, giving an expected chunk size of 16 bytes.
Since the anchors are based on the file contents instead of fixed
sized offsets, an insertion or deletion will only affect at most two
chunks and leave the remaining chunks in the file unchanged.   We note
that our window and chunk sizes are considerably smaller than those
used in LBFS because our goal is to identify tokens that appear across
a file's lifetime, not to optimize the use of network bandwidth.
Thus, the chunk size is chosen to be on the order of the expected
storage required for a single configuration setting.

We call the contents of each chunk the {\em chunk value} and compute
the frequency of each chunk value over all versions of a file.  
%A file
%with high similarity will exhibit only a few unique chunk values, and
%each chunk value will appear in a large portion of its versions.  On
%the other hand, a file with low similarity will have many unique chunk
%values, and each will appear in only a few versions.  -\section{Contributions}

A file with high similarity will have many chunk values that appear in
a large portion of its versions. Conversely, a file with low
similarity will have many chunk values that appear in only a few of
its versions. To assign a numeric score to a file's similarity across
versions we compute the following ratio:
\begin{equation}
Similarity = \displaystyle\frac{\sum_{i=0}^n c_i}{v \times n}
\label{eqn:sim}
\end{equation}
where $n$ is the average number of unique chunks in each version of
the file, $c_i$ is the number of occurrences of the $i$th chunk when
the chunks are sorted in descending order by frequency of occurrence across the different versions,
and $v$ is the number of versions of the file recorded.  Intuitively,
similarity is the ratio between the sum of the occurrences of the $n$
most frequent chunks in the file, over a similar sized file that does
not change at all -- i.e. each chunk appears in every version.  To
illustrate, Figure~\ref{fig:sim} compares the chunk distribution for a
configuration file, {\em prefs.js} and a non-configuration file, {\em
session.rdf}.  A large proportion of chunks in {\em prefs.js} appear
in every version of the file, so the rectangle representing $v \times
n$ is almost entirely covered (96\%) by the histogram.  On the other
hand, {\em session.rdf} experiences much lower similarity among
its versions, and has a long tail of low frequency chunks extending
beyond the shaded rectangle.  As a result, the histogram covers a much
smaller portion of the rectangle (55\%).

%\subsection{Significance Function}

%When the user is trying to recover from a misconfiguration, SAIC can
%help indicate which of the files the failing application is accessing
%contain configuration state.  In addition, to speed up recovery,
%SAIC offers a {\em significance function}, which finds versions that
%contain changes that are likely to affect portions of the file that
%hold configuration state.  The intuition behind this function follows
%from our insight that configuration state changes slowly as compared
%to other application state.  Thus, the significance function looks for
%versions that either introduce a long-lived chunk or remove a
%long-lived chunk.  Currently, SAIC defines a long-lived chunks as a
%chunk that appears in more than 50\% of the versions of a file used to
%compute the similarity metric.

%SAIC also displays the contents of newly added or removed long-lived
%chunks as well as the chunk before and afterwards. We have found this
%to be a useful debugging feature for ASCII configuration files as the
%chunk values often do provide hints as to what configuration change
%was made.

%SAIC defines a long-lived chunk as any chunk that
%is contiguously present for more than 25\% of the files versions.

\section{Implementation}
\label{section:implementation}

\subsection{Computing Similarity}
\label{sec:computing}

Our prototype consists of a tool to compute similarity of a file's
versions and a versioning file system, which is implemented by the
redo logs maintained by the kernel module that we used in
Section~\ref{sec:problem:recovery} to collect our user traces.  When
running in regular system operation mode, it first applies the three
filters to see which file's similarity should be computed, and assigns
a similarity score of zero to those files that do not pass the
filters.  For each file that is not removed by the  filters, the
similarity computation tool extracts each version of the file from the
redo log maintained by the versioning file system and computes
overlapping Rabin fingerprints, which it uses to select anchors to
delimit file chunks.  Because the chunk values have varying lengths, a
hash function is applied to them and the result is then inserted into
a hash table that maintains a running count of how many times a chunk
value has appeared.  This process is repeated for each file version,
thus computing how many times each chunk values appears across all
versions of the file.  If a chunk value appears multiple times in any
particular file version, it is only counted once so that the count of
any hash value cannot exceed the number of file versions.  Finally,
these counts are used as described by Equation~\ref{eqn:sim} and the
score is assigned to the file.  Since configuration files do not
become non-configuration files or vice-versa, SAIC only runs the
tool once per file, thus keeping the performance requirements of
SAIC low.   Once a file is assigned a score it maintains that score
until it is deleted from the file system.

\subsection{Triggering Similarity Measurements}
\label{sec:trigger}

\begin{figure}[tb]
\begin{algorithmic}[1]
\Function{Trigger Similarity}{}
\State $i \gets 1$
\State $v \gets$ next\_version() \Comment{Get the first version}
\While{$i < M$}
\If{more\_versions()}
\State $x \gets$ next\_version() 
\Else
\State {\bf return }{NULL} \Comment{Not enough versions yet}
\EndIf

\If{timestamp($x$) $-$ timestamp($v$) $ > T$}
\State $v \gets x$
\State $i \gets i + 1$
\EndIf

\EndWhile

\State {\bf return }{$v$}

\EndFunction
\end{algorithmic}
\caption{Function for determining when to perform a similarity
measurement.  If there are not enough versions to get $M$ sampling
periods of length $T$, the function returns NULL.}\label{algo:sampling}
\end{figure}

While SAIC's similarity algorithm is effective at identifying
configuration files, an important requirement is that SAIC must be
able to provide online measurements to a versioning file system
efficiently.  During regular operation, SAIC only invokes the
similarity tool when there are idle cycles available on the processor.
This is done to minimize the impact of SAIC on the usability of the
user's machine.  While the number of cycles allocated to SAIC will
depend on the load on the machine, we assume that SAIC must have
very meager resource requirements, and must be tolerant to bursty
resource availability.  As a result, SAIC cannot take a measurement
after every file modification.  Despite this, we expect SAIC to
produce accurate measurements in a timely manner so that the
versioning file system can make informed decisions about which files
to discard when it hits its space quota.  Thus, to make the best use
of the few cycles allocated to it, SAIC must trigger a similarity
measurement only on files that have had enough activity to produce an
accurate similarity score.

%In addition, since configuration files and non-configuration
%files are exclusive to each other, there should be no reason to
%measure a file again once an accurate score has been assigned to it.

%As a result, SAIC  must determine when to take a similarity
%measurement.

%When used to analyze a file system trace offline, it is relatively
%straightforward to run SAIC on every file version in the entire
%trace.  However, when run online, 

Finding the correct time to take such a measurement is an important
factor in the effectiveness of SAIC.  On the one hand, waiting too
long to take a measurement carries several risks.  First, the
versioning file system may mistakenly continue to version a
non-configuration file and instead discard a configuration file's
versions because SAIC has not given it the information necessary to
make a correct decision.  Second, SAIC's run time increases with the
number of file versions it must process.  Finally, all files that are
modified exhibit decreasing similarity over time, but the rate of
similarity decrease in configuration files is much lower than in
non-configuration files.  However, waiting too long to compute a
similarity measurement of a configuration file will give it an
artificially low score, causing it to be incorrectly discarded.

On the other hand, like all statistical analysis, SAIC can produce
noisy results if run over too little data.  As a result, taking a
measurement too early can give an inaccurate score due transient
events on the file.  For example, some configuration files experience
a large number of changes when they are first created, and then
eventually settle into a more representative pattern of low activity.
Taking a similarity measurement during the transient at the beginning
would produce an artificially low similarity score.  Thus, it is
critical that SAIC picks the correct time to take a measurement of a
file and that the time of measurement be normalized across all files.

Our initial thought was that there exists a fixed number of versions
after which a similarity measurement would return an accurate result.
This turned out to be false because the rate at which new versions are
created varies greatly from file to file.  As a result, using a fixed
number of versions would give files that are modified very often an
artificially high similarity score because no task-dependent or
time-dependent events will have occurred between versions created
extremely close together.   In other words, when files are rapidly
modified, the changes between each version are small, making the file
have an apparently high similarity.

This observation made us realize that an algorithm to pick the correct
measurement time must take into account not only the number of
versions of a file that exist, but also the time over which the
versions are created.  Such an algorithm should discount versions
created close together since it is unlikely that there was an
intervening task-dependent or time-dependent event and reward versions
created farther apart in time.  To do this, we defer computing a
similarity measurement until a file has had $M$ modifications, each of
which are at least $T$ hours apart.  In other words, there must be at
least $M$ non-overlapping time periods of length $T$ that contain at
least one new file version.  

The point at which SAIC computes similarity for a file is called the {\em trigger point} and is found
by using the {\em trigger function} described in
Figure~\ref{algo:sampling}.  We call $T$ the sampling period and $M$
the sample length.  The trigger function will return the version at
which a measurement should be taken if enough versions of the file
have been generated in the versioning file system.  If there have not
been enough versions, then the function returns NULL.   Ideally, the
sample period should be long enough to contain one task-dependent or
time-dependent event.  Thus, using file versions across a number of
sample periods will contain enough task-dependent changes that
non-configuration files that contain task-dependent state will
produce a lower similarity score than configuration files that contain
mainly configuration state.  In practice, we have found that a sample
period of 3 hours and a sample length of 12 works fairly well.

%We note that a file may be used in recovery mode before it has hit its
%trigger point, and thus would have no similarity score associated with
%it.  
In recovery mode, the user may need the  score of a file that has not
reached its trigger point yet, and thus has not had a similarity
measurement performed on it.  In this case, SAIC computes the
similarity on all versions up to the point that the recovery is taking
place and uses that similarity score instead.  This score is only used
temporarily during recovery as a best approximation of what the
trigger point score would be.  It is discarded after the user returns
to regular operation and the file must still reach its trigger point
before SAIC will assign a permanent score to it.

\begin{comment}
\begin{figure}[tb]
\center
\footnotesize
\begin{boxedverbatim}
i = 0;

/* Get the first version */
v = get_next_version();

while (i < M) {
  x = get_next_version();

  /* Are we at the end of the redo log? */
  if (!is_valid(x)) {
    break;
  }

  /* Check if timestamp on this version is 
     more than one time quantum away from the 
     previous version. */
  if (time_of(x) - T < time_of(v)) {
    v = x;
    i++
  }  
}

/* check if we found enough versions.  
   If so, return the last version. */
if (i == NUM_OF_QUANTA) return v;
else return NULL;

\end{boxedverbatim}
\normalsize
\caption{Pseudo-code describing the algorithm used to trigger a
similarity measurement.  $T$ represents the sampling period and $M$
represents the sampling length.}\label{fig:sampling}
\end{figure}
\end{comment}

\subsection{Improving Performance}
\label{sec:sampling}

%Since most of the time is spent
%extracting file versions, computing Rabin fingerprints, computing
%hashes of chunk values and inserting them into the hash table, 

During regular operation, we found that some files took a long time to
compute their similarity scores.  However, in this mode, SAIC must
produce similarity scores quickly and without the need for a large
amount of resources.  There are a number of operations that scale with
the aggregate size of the file versions used in the computation.  For
example, the time to extract file versions, the number of Rabin
fingerprints to compute and the number of hash computations all
increase with the number of versions and the size of the file.
Another source of overhead is the the size of the chunk value hash
table.  For a file with extremely low similarity and large aggregate
size of file versions, the number of unique chunk values can become
very large.  This added memory pressure results in thrashing on the
machine and greatly degrades performance.

%However, this did not completely account for the long computation
%times we experienced for some files.  It turns out that another
%important factor to the run-time of SAIC's similarity measurement is
%.   .   For example, for {\em
%VMware-server-2.0.0-122956.i386.tar.gz}, there are roughly 35327990
%unique chunk values, which would require 3GB of memory to store the
%hash table for.   

%Low similarity files often have a long tailed distribution of chunks,
%with many chunk values appears very few times.  Since the similarity
%computation only considers the $n$ most frequentlly occuring chunks,
%these 

As a result, the overhead of the similarity computation is determined by
the file's similarity, the size of the file and the number of versions
used in the similarity computation.  Since we can't control the size
or similarity of the file, we must reduce the number of versions used
to make these files practical to measure.  We employ the sampling
periods used in the triggering function and use only the first sample
of each sample period for the similarity computation, discarding the
rest of the versions within the sampling period.  In this way, the
number of versions used in the similarity computation is bounded by
$M$, which is 12 in our prototype.  The intuition behind why this
approximation works is that even if the rate of change of a file is
significantly faster than the sampling period, the samples will still
capture the cumulative changes and produce a similarity score close to
the one that would have been achieved by sampling all versions.  The
only case where the scores may significantly differ is if a file was
being modified in a cyclic way, and eventually returned back to a
state in a previous version.  We did not observe any files that
displayed this behavior.  We will compare the scores derived using
sampling and using all versions in Section~\ref{section:evaluations}.

%% file: evaluation.tex
\section{Evaluation}
\label{section:evaluations}

%& VMware & 264 & & 10 & 17489/17556(99.62\%) & 3/67 \\ \hline
%\multirow{5}{*}{2} & Firefox & 3488 &  & 45 & 3029332/3123853(96.97\%) & 18/94521 \\ 
%& GNOME & 958 & & 144 & 229655/229926(99.88\%) & 14/271 \\
%& Flash & 26 & & 21 & 87/87 (100\%) & 0/0 \\ 
%	& JEdit & 178 & & 7 & 2774/2787(99.53\%) & 3/13 \\
%& Amarok & 3143 & & 15 & 29199/29443(99.17\%) & 5/244 \\ \hline
%\multirow{4}{*}{N/A} & Acrobat & & & & & \\ 

\subsection{Identifying Configuration Files}
%\label{sec:eval}

We now evaluate SAIC's ability to identify configuration files.
Several metrics are of importance for SAIC.  One such metric is the
percentage of non-configuration file versions SAIC eliminates.  This
number tells us how many fewer file rollbacks a user must perform when
trying to find the most recently working application configuration.
As a result, it can be used as an indicator of how much effort SAIC
will save the user during recovery.  Another useful metric is the
amount of versioning space SAIC has saved.  This saves the
versioning file system from versioning files unnecessarily, thus
either reducing the space overhead of the versioning file system or
allowing it to maintain longer file histories.  Finally, we would like
to evaluate SAIC's accuracy by measuring its true positive and
false positives when identifying files. We define a true positive as
when SAIC correctly identifies a configuration file, and a false
positive as when SAIC incorrectly identifies a non-configuration
file as a configuration file.  We do not evaluate the overhead of the
file system versioning in SAIC because we have not made any effort
to optimize the kernel module that was performing the file versioning.  The overheads of optimized versioning file systems have
been well studied in the literature~\cite{ElephantFS, versionfs,
wayback}.

%several of SAIC's characteristics.  First, we
%evaluate the effectiveness of the trigger function described in
%Section along two metrics: how well it is able to differentiate
%configuration files from non-configuration files and how many versions
%it saves the versioning file system form unecessarily maintaining by
%delivering scores earlier in our two user traces.  Then, we will then
%evaluate the effectiveness of computing similarity using only one
%version from each sampling period.  The main reason for doing this is
%to reduce SAIC's resource requirements, but we also ensure that it
%does not impact SAIC's accuracy.  Finally, we evaluate SAIC's
%significance function to see if it can effectively reduce the number
%of files versions an administrator would have to try to find the more
%recently working application configuration.

We evaluate each of the three implementation options described in
Section~\ref{section:implementation}.  First, we evaluate the accuracy of SAIC
when applied to all file versions in our two traces.  We then compare
the accuracy of this option to SAIC using the trigger function
described in Section~\ref{sec:trigger} and then finally the option using the
sampling performance enhancement described in
Section~\ref{sec:sampling}.

%This will then
%be evaluated against SAIC when only considering versions up to the
%trigger point.  This is more representative of the accuracy one would
%get from SAIC during regular operation.  Finally, we evaluate SAIC
%using both the trigger point and the performance enhancement achieved
%by sampling to limit the number of versions used by the similarity
%tool

\subsection{Similarity Evaluation}

\begin{table*}[t!]
\center
\begin{tabular}{|l|l|r|r|r|r|r|} \hline
\multirow{2}{*}{\bf Trace} &
\multirow{2}{*}{\bf Application} &
\multirow{2}{*}{\bf Versioned Files} &
\multicolumn{3}{|c|}{\bf Filters} &
\multicolumn{1}{|c|}{\bf Passed} \\ \cline{4-6}
& & &
\multicolumn{1}{|c|}{\bf Persistent} &
\multicolumn{1}{|c|}{\bf Read$\rightarrow$Write}&
\multicolumn{1}{|c|}{\bf User file}&
\multicolumn{1}{|c|}{\bf Filter}
\\ \hline \hline
\multirow{4}{*}{1} & Firefox & 507 & 228 & 265 & 1 & 13 \\
& GNOME & 392 & 9 & 363 & 0 & 20\\
& Flash & 17 & 12 & 0 & 0 & 5 \\
& VMware & 189 & 105 & 7 & 71 & 6 \\ \hline
\multirow{5}{*}{2} & Firefox & 568 & 449 & 86 & 3 & 30 \\
& GNOME & 493 & 316 & 151 & 0 & 26 \\
& Flash & 35 & 24 & 5 & 0 & 6 \\
& JEdit & 73 & 12 & 41 & 14 & 6 \\
& Amarok & 89 & 29 & 51 & 4 & 5 \\ \hline
%\multirow{4}{*}{N/A} & Acrobat & & & & & \\
%& RealPlayer & & & & & \\ 
%& Wordpress & & & & & \\ 
%& OpenOffice & & & & & \\ \hline
\end{tabular}
\caption{Filter results.  This table gives the number of files removed
by each of the filters.  The last column indicates how many files
from each application had the similarity computation applied to them.}
\label{tbl:filter}
\end{table*}

%We also generated artifical traces with some additional applications:
%the Adobe Acrobat Reader, RealPlayer Media player, the Wordpress 2.0
%web application and OpenOffice.   These traces were gathered by
%intensively using the applications for about a day or so.  

\myfigwide{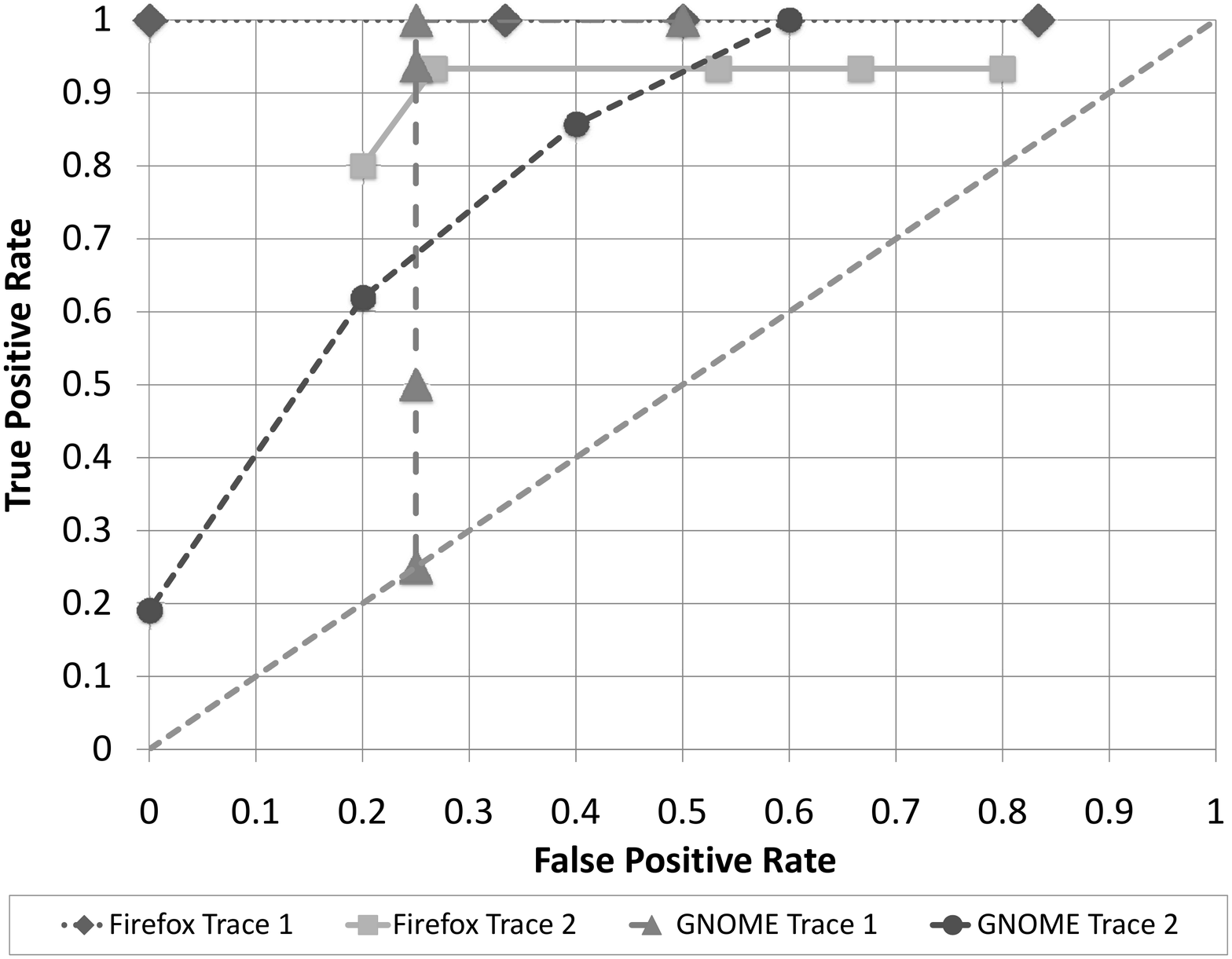}{Receiver Operating Characteristic curve for SAIC's Similarity Metric.
The points from left to right in each curve represent similarity
thresholds of 95\%, 90\%, 80\%, 70\% and 60\%.  We note that Firefox
in Trace 1 actually experiences a 100\% true positive rate at all
thresholds because all Firefox configuration files scored above 95\%
in that trace.}{fig:ROC}{0.75}

We began by using SAIC to measure similarity across all versions of
all files in our trace.  As described in Section~\ref{sec:model},
SAIC first applies a set of filters to remove non-application state
files.  We found that these filters play an important role because a large portion of versioned files are actually removed from
further consideration by these filters.  This is beneficial because
the filters are far more efficient than the similarity tool.
Table~\ref{tbl:filter} shows the number of files filtered out for each
application by each filter.  The passed filter column indicates the
number of application state files that are not removed by the filter,
which will have the similarity calculation applied to them.  We
verified that none of the files removed by the filters were
configuration files.  As we can see, the majority of files modified by
each application actually consist of temporary and non-persistent
files.

%We observed that a common behavior for
%applications was that when they were going to modify a file -- they
%would create a temporary file with the contents of the new file and
%then rename the temporary file over the old file.  

%Thus, even applications like the GNOME session
%manager, which deals mainly with configuration files have a large
%number of temporary files.

Next the tool computes the similarity scores of all the files that
passed the filters. Most configuration files scored greater than the
majority of non-configuration files.  To illustrate,
Figure~\ref{fig:ROC} shows a receiver operating characteristic (ROC)
curve, which shows what the accuracy of SAIC would be at similarity
thresholds of 60\%, 70\%, 80\%, 90\% and 95\%.  Files with scores
above the threshold are identified as configuration files and files
below the threshold are identified as non-configuration files.  The
ROC curve plots the true positive rate versus the false positive rate.
The true positive rate is defined as the number of true positives over
the actual number of configuration files, while the false positive
rate is the number of false positives over the number of
non-configuration files that were not removed by the filters.  For
readability, we only show curves for the four applications which have
the most versioned files.

In reality, SAIC's versioning file system does not use an explicit
threshold to determine whether a file should be versioned or
not. Rather, this threshold is determined implicitly by how much space
is available for versioning -- the versioning file system simply
discards the lowest scoring file when it is low on space.  As a
result, the more storage is allocated, the lower the implied
threshold, although the precise relationship between version storage
and similarity threshold will depend on the amount of versioning
activity and the size of the files currently in the version store.
From the curve we see that as the point moves to the right along
the curves, which means threshold is becoming lower, both the false positive and false negative rates will
rise.  The closer a point is to the upper left corner (100\% true
positives with no false positive) the better the operating point.  The
important thing to note is that the curves are all well above the
``line of no discrimination'' ($x = y$), indicating that SAIC
generally has a higher true positive rate than false positive rate.
By computing the Euclidean distance between each of the operating
points and the upper left corner, we determine that the optimal
operating point for the files in our trace is a similarity threshold
of 82\%.  90\% of all configuration have similarity scores above this
threshold and the false positive rate at this operation point is 35\%.
At this point SAIC will eliminate 85\% of all non-configuration file
versions and save 77\% of versioning space in the redo logs.  We note
that with a threshold of 55\% SAIC is capable of identifying all
configuration files while still correctly eliminating 30\% of
non-configuration files.  At this operating point, SAIC suffers only
33 false positives across both traces and is able to eliminate roughly
66\% of non-configuration file versions from its redo log.

\begin{table*}[t!]
\center
\begin{tabular}{|l|l|r|r|r|r|r|r|r|} \hline
\multicolumn{1}{|c|}{\bf Trace} &
\multicolumn{1}{|c|}{\bf Application} &
\multicolumn{1}{|c|}{\bf TP} &
\multicolumn{1}{|c|}{\bf FN} &
\multicolumn{1}{|c|}{\bf FP} &
\multicolumn{1}{|c|}{\bf Versions eliminated} &
\multicolumn{1}{|c|}{\bf (\%)} &
\multicolumn{1}{|c|}{\bf Space Saved (MB)} &
\multicolumn{1}{|c|}{\bf (\%)}
\\ \hline \hline
\multirow{4}{*}{1} & Firefox & 7 & 0 & 3 & 258505/289356 & 89.3 &
1025/1151 & 89.1\\ 
& GNOME &  15 & 1 & 1 & 1737/1743 & 99.7 & 62.8/62.8 & 100\\
& Flash &  1 & 0 & 0 & 123/123 & 100 & 0.0138/0.0138 & 100\\
& VMware &  4 & 2 & 0 & 2766/2766 & 100 & 32.0/32.0 & 100 \\ \hline
\multirow{5}{*}{2} & Firefox &  14 & 1 & 8 & 422066/516984 & 81.6 &
4733/5060 & 93.5\\ 
& GNOME &  18 & 3 & 2 & 17056/17090 & 99.8 & 1267/1267 & 100\\ 
& Flash & 1 & 0 & 2 & 1190/1980 & 60.1 & 0.784/1.46 & 53.7 \\ 
& JEdit &  2 & 0 & 1 & 4940/4956 & 99.7 & 51.5/51.5 & 99.8\\ 
& Amarok &  3 & 0 & 1 & 4735/4742 & 99.9 & 194/194 & 100 \\ \hline
%\multirow{4}{*}{N/A} & Acrobat & & & & & \\ 
%& RealPlayer & & & & & \\ 
%& Wordpress & & & & & \\ 
%& OpenOffice & & & & & \\ \hline
\end{tabular}
\caption{Performance of SAIC over all versions in each trace. Here we
use a similarity threshold of 80\% to differentiate configuration
files from non-configuration files.  TP denotes the number of true
positives, FN, the number of false negatives (configuration files that
were not correctly identified) and FP, the number of false positives.
The Versions Eliminated column gives the number of  non-configuration
file versions eliminated over the total number of non-configuration
file versions.  Similarly, the Space Saved column gives the amount of
versioning space in MB that was eliminated over the total potential
space that is used to version non-configuration files.}
\label{tbl:entire_trace}
\end{table*}

To evaluate the number of versions SAIC can eliminate and space
SAIC can save, we will use the 80\% similarity threshold as our
assumed operating point.  Table~\ref{tbl:entire_trace} gives the
number of true positives, the number of false positives and the number
of false negatives (configuration files wrongly identified as
non-configuration files).  To evaluate the amount of effort the user
saves during rollback, we provide the number of non-configuration file
versions eliminated over the total number of non-configuration file
versions.  The amount of space saved is given by the amount of
non-configuration file space saved in the redo logs over the total
space used to version non-configuration files.  Any of the versions
eliminated or space that was not saved is a result of false positives.
We can see that SAIC will help the user tremendously during recovery
for a majority of the applications by correctly identifying all
non-configuration files and eliminating them from consideration during
recovery, leaving only actual configuration files for the user to try
to recover with.  Similarly, nearly all the versioning space that
would have been wasted on non-configuration files is saved.  The only
exception was the Flash application in Trace 2.  Here, the vast
majority of all versions not eliminated were the result of a single
false positive file, {\em clearspring.sol}, which had 787 versions and
a similarity score of 87\%.

Only Firefox had significant numbers of false positives.  Of these,
two files, {\em cookies.sqlite} and {\em places.sqlite}, account for
99.9\% of the misidentified versions, which were not eliminated as a
result.  These files have a common characteristic -- they are
histories of user activity and behave as a slowly changing cache for
that activity.  As a result, even though the entries in these files
are task-dependent, some can remain in the file for a long time either
because they are used frequently or there is insufficient activity in
the application to evict them.  We note that {\em
urlclassifier3.sqlite} and {\em pluginreg.dat}, the two problematic
configuration files described in Appendix~\ref{sec:appendix:apps},
were both correctly identified by SAIC.

Both false negatives in VMware were in a file that VMware uses to
store the user's favorite VMs.  While this is configuration state,
VMware also uses this file to store a large amount of task-dependent
state, causing the file to have low similarity.  We note that, with
the exception of {\em clearspring.sol}, the false positive in the
Flash application, many of the false positives have negligible impact
on the number of versions and the amount of space saved.  This is
because the false positives often turned out to be short files with
few versions.  We will discuss the reasons for this in
Section~\ref{sec:eval:discussion}.

%In all cases, SAIC eliminates over {\bf xx} of the
%non-configuration file versions and {\bf xx} of the version storage
%space for non-configuration files.  Only Firefox had significant
%numbers of false positives.  Two of these files, {\em
%formhistory.sqlite} and {\em places.sqlite}, account for 99.9\% of the
%misidentified versions.  These files have a common characteristic --
%they are histories of user activity and behave as a slowly changing
%cache for that activity.  As a result, even though the entries in
%these files are task-dependent, some can remain in the file for a long
%time either because they are used frequently or there is insufficient
%activity in the application to evict them.  The remainder of the false
%positives were of small files with relatively few versions, allowing
%SAIC to eliminate nearly all versions and storage for
%non-configuration files.

\begin{table*}[t!]
\center
\begin{tabular}{|l|l|r|r|r|r|r|r|} \hline
\multirow{2}{*}{\bf Trace} &
%\multirow{2}{*}{\multicolumn{1}{|c|}{\bf Application}} &
\multirow{2}{*}{\bf Application} &
\multicolumn{2}{|c|}{\bf All Versions} &
\multicolumn{2}{|c|}{\bf Trigger} &
\multicolumn{2}{|c|}{\bf Sampling} \\ \cline{3-8}
& &
\multicolumn{1}{|c|}{\bf TP} & 
\multicolumn{1}{|c|}{\bf FP} & 
\multicolumn{1}{|c|}{\bf TP} & 
\multicolumn{1}{|c|}{\bf FP} & 
\multicolumn{1}{|c|}{\bf TP} & 
\multicolumn{1}{|c|}{\bf FP} \\ \hline \hline
\multirow{4}{*}{1} & Firefox & 7/7 & 3/6& 7/7 & {\bf 4/6} & 7/7 & {\bf
4/6} \\ 
& GNOME & 15/16 & 1/4 & 15/16 & 1/4 & 15/16 & 1/4 \\ 
& Flash & 1/1 & 0/4 & 1/1 & 0/4 & 1/1 & 0/4 \\ 
& VMware & 4/6 & 0/0 & 4/6&0/0 & 4/6 & 0/0 \\ \hline
\multirow{4}{*}{2} & Firefox & 14/15 & 8/15 & 14/15& {\bf 9/15} &
14/15 & {\bf 8/15} \\ 
& GNOME & 18/21 & 2/5& {\bf 19/21} & 2/5 & {\bf 21/21}  & 2/5 \\ 
& Flash & 1/1 & 2/5 & 1/1 & 2/5 & 1/1 & 2/5 \\ 
& JEdit & 2/2 & 0/4 & 2/2 & 0/4 & 2/2 & 0/4\\
& Amarok & 3/3 & 0/2 & 3/3 & 0/2 & 3/3 & 0/2 \\ \hline
%\multirow{4}{*}{N/A} & Acrobat & & & & & & \\ 
%& RealPlayer & & & & & & \\ & Wordpress & & & & & & \\ & OpenOffice &
%& & & & & \\ \hline
\end{tabular}

\caption{Evaluation of Trigger Function and Sampling.  We show the
effects of both mechanisms on the accuracy of SAIC.  In most cases,
these optimizations had no effects on the accuracy.  In cases where
there was an effect, the change is highlighted in bold.}\label{tbl:trigger}
\end{table*}

\subsection{Trigger Function Evaluation}

To evaluate the effectiveness of the trigger function, we modify our
similarity measurement tool to only compute similarity for file
versions up to the file's trigger point and compare its accuracy at
identifying configuration files with the original tool that uses all
versions available in the trace.  The results are given in
Table~\ref{tbl:trigger}.  When a file's trigger point does not occur
before the end of the trace, we use all the versions gathered in the
trace.  This simulates SAIC's recovery mode behavior, which is to
use all available versions for files that have not reached their
trigger points.  On average, using a trigger point increases the
similarity score of a file by 2$\pm$6\%.  This was expected since most
files have a decreasing similarity score over time, so taking a
measurement earlier will increase a file's similarity score on
average. As we can see from the results, this has a negligible effect
on the effectiveness of SAIC.  In two cases, a non-configuration
file had its similarity score increased to exceed the 80\% threshold
and become a false positive.  In both these cases, the file was the
user's Firefox cookie jar, which increased from 79\% to 98\% in Trace
1 and from 73\% to 85\% in Trace 2. Increasing similarity scores can
also have positive results --  the similarity score of one GNOME
configuration file was increased from 79\% to 83\% and went from a false
negative to a true positive.

The benefit of the trigger function is that SAIC can provide
results to the versioning file system much earlier.  Of the 96239 
versioned files in both traces, 149 files reached their trigger point
before the end of the trace.  The cumulative distribution of times to
reach the trigger point for both traces is given in
Figure~\ref{fig:trigger_trace1} and Figure~\ref{fig:trigger_trace2}.
In Trace 2, which had more activity, many more files were able to reach
their trigger points with several days of activity while in Trace 1,
which had less activity, half the files took 18 or more days.  While
the time for the trigger point to occur does depend heavily on how
much the application that accesses the file is used, we take these
numbers to show that SAIC can get accurate measurements for a fair
number of files within the scope of several weeks.

%We also evaluated how much earlier in the traces the trigger function
%can produce measurements, as opposed to versioning all the way to the
%end of the trace.  The results are given as histograms in
%Figure~\ref{blah} showing the distribution of time savings as a
%percentage of the length of the trace.

\myfig{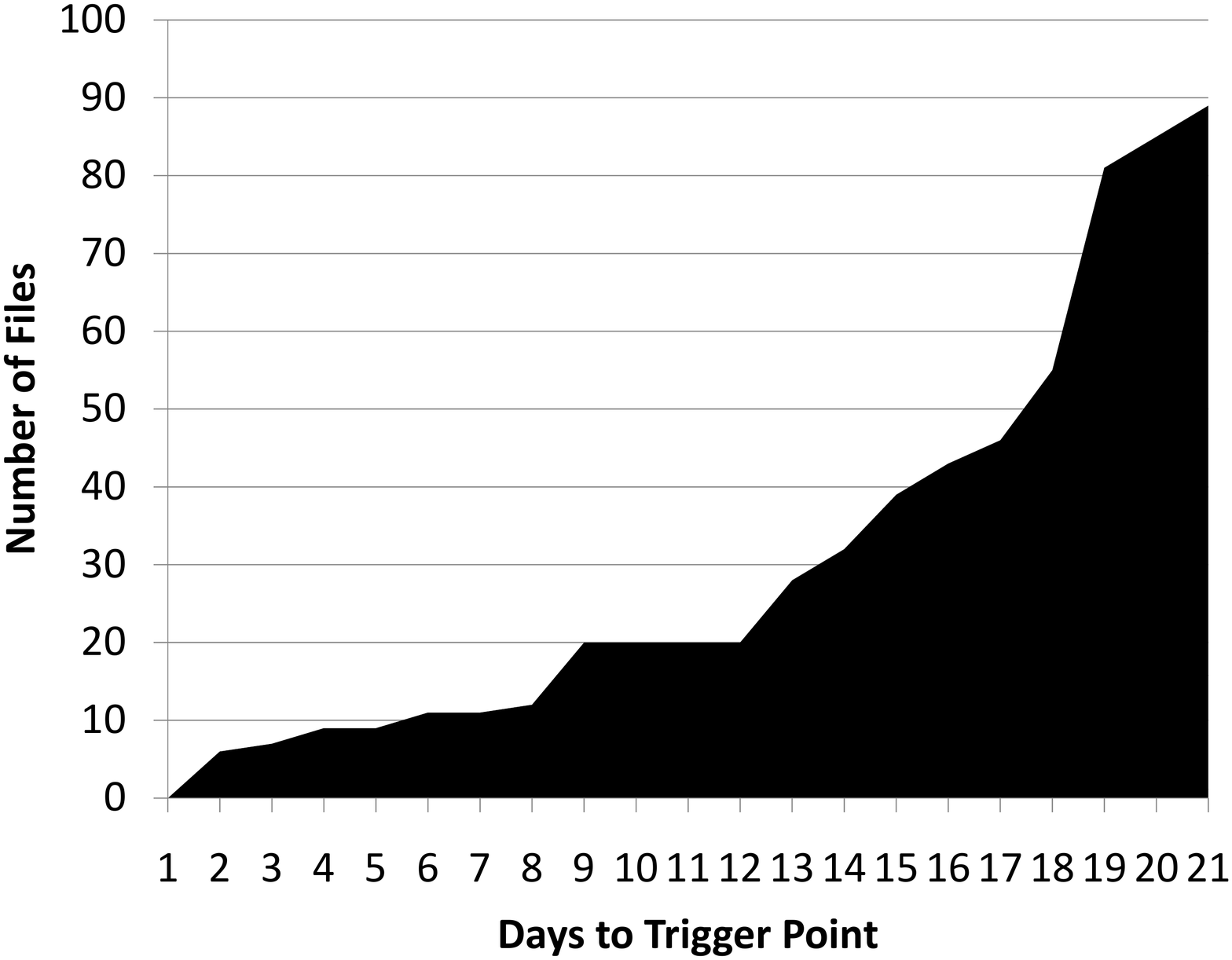}{Cumulative distribution of time for files to reach
their trigger points in Trace 1.}{fig:trigger_trace1}{1.0}

\myfig{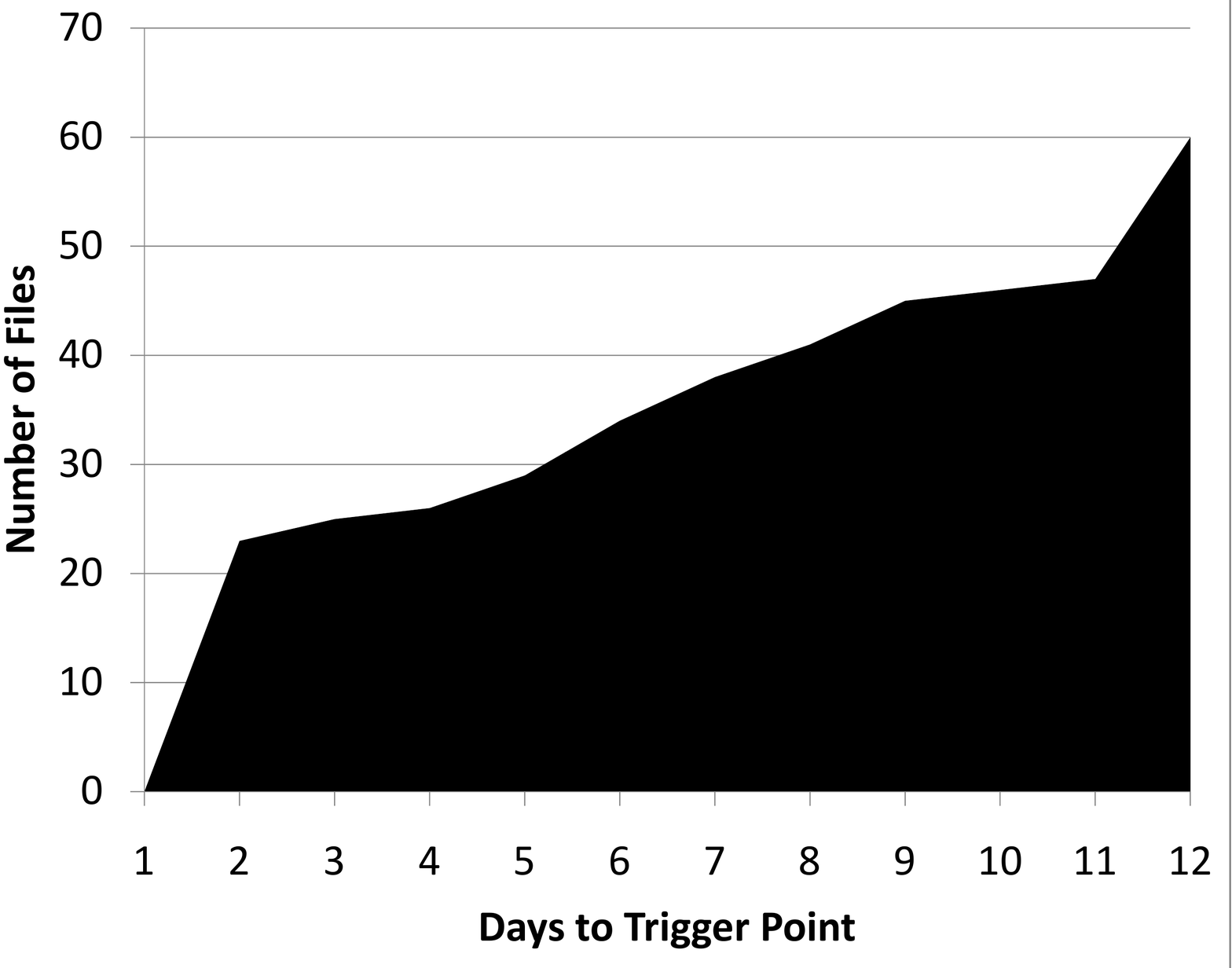}{Cumulative distribution of time for files to reach
their trigger points in Trace 2.}{fig:trigger_trace2}{1.0}

\subsection{Sampling Evaluation}

While sampling reduces the resource requirements of SAIC, we also
wanted to measure how sampling perturbs SAIC's similarity score as
compared to computing the score with all versions up to the trigger
point.  Table~\ref{tbl:trigger} also shows the differences in
performance when we configure SAIC to only use one file version from
each sampling period up to the trigger point versus using all versions
up to the trigger point.  Sampling perturbs the similarity score of a
file by 2$\pm$7\% on average and as a result has essentially no
measurable affect on the accuracy.  Anecdotally, two configuration
files, each of which had a similarity score of 78\%-79\% when computed
using all samples up to the trigger point, had their similarity scores
increased to 80\%-84\% by sampling.   One false positive introduced by
using the trigger function also decreased below the threshold, becoming a
true negative again.  Using sampling periods decreases the time
required to analyze the files by 86\% overall.  In particular, some of
the larger files saw the time to compute their similarity score drop
by over 90\%.

\input{discussion}

%% file: discussion.tex
\section{Discussion}
\label{sec:eval:discussion}

Because SAIC's similarity metric relies on heuristics, it can make
mistakes when these heuristics fail.  As we have seen, some
non-configuration files may also exhibit high similarity by
coincidence -- for example they may record user histories.  However,
the impact of such errors is lessened by two properties of SAIC.
First, SAIC is more likely to classify a non-configuration file as a
configuration file than the reverse. This is because while there is
non-configuration state that can change slowly, we have not observed
any configuration state that changes frequently.  We believe that such
a situation would only arise if a user changed configurations as often
as she changed tasks, or a bug in the application caused
configurations to spontaneously change on their own -- both are
exceptional circumstances.  Hence, a misidentification will cause more
than necessary disk space to be used and extra effort during recovery,
but will not result in important file versions getting
discarded. Second, because SAIC relies on identifying common chunks
in file contents over time, it can misclassify files when there
have not been enough versions, or when the file is very small. In such
cases, results can have quite a bit of error associated with them.
Fortunately, it is not expensive to version small files with few
versions, so one solution may be to augment the file system's
versioning policy to retain such files regardless of their similarity
score.

%% file: related.tex
\section{Related Work}\label{section:background}
This section presents the most important research related to this thesis. Firstly, we highlight the difference of our work with other work in configuration management. Secondly, we explore work in system recovery and versioning file systems. Finally, we present work that used Rabin fingerprint.
\subsection{Configuration Management}
There has been a lot of interest in configuration management. Recent work \cite{Nagaraja2004, Oliveira2006, Oppenheimer2003, Jiang2009} shows that configuration problem is a significant cause of system faults. Broadly speaking, research in this field can be classified into three categories: providing aids to the users to fix configuration problems, validating user actions that change configurations, automating some error-prone tasks of configuration management.

Much of this work \cite{chronus, autobash, Mona2008, peerpressure, strider, snitch, netprints, Yuan2006} uses computers to aid the process of configuration problem troubleshooting. Chronus \cite{chronus} searches for the time when system configuration transitioned from a working state to a non-working configuration state. During the diagnosis process, Chronus loads and runs historical system snapshots in a virtual machine monitor, and tests whether a historical system snapshot works correctly by executing user-provided software probes. To keep a history of the system snapshots, Chronus logs every change to the disk in a time-traveling disk. In addition, it uses binary search to speed up the search of the state transition point.

SAIC is inspired by the idea of searching for historical working state of Chronus. However, it differs from Chronus in two ways. First, a major problem of the Chronus method is that it can not differentiate configuration-related disk block changes with user data-related disk block changes. Thus it can inadvertently cause user data loss with its indiscriminating disk block roll back. Chronus solved the problem by rolling back only configuration-related file changes. Second, binary search can not guarantee to find the latest working state if there are multiple working state to non-working state transitions. SAIC ensures that the latest working state is found by using linear search. 

Autobash \cite{autobash} and its successor \cite{Mona2008} help a user to locate existing solutions to configuration problems. They maintain a database of user actions to known configuration problems and uses a trial-and-fail approach to test candidate solutions. To ensure that an incorrect solution does not leave persistent changes to the system, they apply each candidate solution in a speculative environment. Any persistent changes made by a solution can be undone if the solution is not appropriate for the configuration problem that the user is troubleshooting. Like Chronus, they execute user-specified software predicates to automatically test whether the configuration problem is solved or not. In addition, they leverage causality-tracking and analysis to decrease the time for finding and testing solutions. Most recently, Su et al. \cite{Su2009} further enhances Autobash by using heuristics algorithms to automatically find user actions that are used to test system states as software predicates.

While Autobash allows a user to learn from other users that experiences the same or similar configuration problems, it can not solve the problems that do not have existing solutions. SAIC is able to fix these kind of problems since it can find out the historical state when the user's system was working.

PeerPressue \cite{peerpressure} and its predecessor, Strider \cite{strider} uses statistical analysis to find the root causes of misconfigurations. They maintain a database that stores the system state of a large body of computers. To diagnose a configuration problem, they first trace the faulty application and use causality-tracking to find suspect Windows registry entries that may cause the configuration problem. Then they retrieve the same set of entries from the system state of other computers from the database. Finally, all these entries are used in Bayesian statistical estimations to calculate the probability for each entry to be the root cause. 

%Like other statistical approaches, the success of PeerPressure in finding the correct root cause misconfigurations for an application requires that the number of correct configurations of the same application that runs on other computers significantly outweighs the number of incorrect configurations of the same application that runs on other computers. Although this requirement can often be met, there are also many cases it can not.

Like other statistical approaches, the success of PeerPressure in finding the correct root causes of misconfigurations for an application requires the use of a large set of system states from computers that runs the same application. This requirement may be feasible for enterprise users that have a large number of installations of computers that run the same set of applications. But it is difficult for an individual user to collect this kind of information from other people's computers. A software vendor is also unlikely to do so due to privacy and other issues. SAIC avoids this kind of problems since it does not need information from other users' computers.

Some research \cite{Nagaraja2004, Oliveira2006} provide frameworks to validate configuration changes before they are put in effect on production systems. Nagaraja et al. \cite{Nagaraja2004} explored the nature of operator mistakes with operator experiments on system maintenance and troubleshooting tasks. They implemented a prototype validation framework with two alternative techniques: trace-based validation and replica-based validation. Oliveria et al. \cite{Oliveira2006} conducted a survey on a group of database administrators to study database configuration problems. They introduced a new model-based validation technique in its validation framework, which is based on \cite{Nagaraja2004}. Both studies indicate that operator mistakes are significant source of system faults. The experiments with these validation frameworks illustrate that they can effectively detect and prevent some operator mistakes, but they miss a considerable number of operator mistakes due to the complexity of the interactions between the operator and the system.

Others \cite{Zheng2007, Anderson2002, Anderson2003, Burgess1995} have proposed to use computers to accomplish some of the most tedious and error-prone tasks of configuration management. These systems allow the user to specify the rules that the system will follow to automatically generate configuration files. They usually provide the user with some high-level language directives or templates that define how the configuration files should be generated under certain circumstances. Although they relieve some of the burden of configuration management from the users, they still rely on the user to supply error-free rules, which can become complicated since there are numerous different kinds of circumstances in which a configuration file needs to be changed. 

%\textbf{mirage? performance misconfiguration? decision-tree based?}

\subsection{System Recovery}
Computer systems can fail for many different reasons \cite{Gray1986}. It has been reported that software bugs and operator mistakes are the most significant sources of system failures  \cite{Gray1990, Jiang2009, Nagaraja2004, Oliveira2006, Marcus2000, Baker92}. Despite the fact that many mechanisms have been proposed to detect software bugs or to identify operator mistakes, many bugs or operator mistakes still bypass the detections and cause system failures, largely due to the sophisticated nature of these issues. Consequently, we believe a more pragmatic solution is to enable computer systems to automatically recover from failures. We present here some of the primary work in this field.

Previous work has used reboot or restart techniques, including whole program restart \cite{Sullivan1991, Gray1986}, microreboot \cite{Candea2004}, and software rejuvenation \cite{Vaidyanathan2005, Kolettis1995} to bring the system to a clean state after system failures. Whole software restart can cause substantial software down time, although it is the most effective way to recover from some system failures. Microreboot reduces restart time by recursively restarting the smallest recoverable component, a software module for example, to the entire software. In many cases, a software module restart can restore the availability of the system and it incurs much less time than a entire software restart. However, being able to microreboot requires the software to be built as a collection of "recoverable" components. While legacy applications can not enjoy this technique, many complicated applications will be difficult or nearly impossible to be designed this way.

Checkpoint and recovery \cite{Feng2005, Laadan2007, assure, King2005, Bressoud1996, zap, Lenharth2009} is another important failure recovery technique. It reduces system down time by periodically taking a checkpoint of the system state and restarting the system from a prior checkpoint. A checkpoint consists of at least the memory state of one or more applications \cite{Elnozahy2002}. Many times it also includes a partial or full file system snapshot. The checkpoints can be stored to disk, non-volatile memory, or remote systems. 

Rx \cite{Feng2005} periodically uses light-weight checkpoint techniques that store a snapshot of an application in main memory and keep a copy of each accessed files of the application. When a software failure is detected in the application, Rx rolls back the application to a recent checkpoint and re-executes the application in a modified environment. This mechanism is able to recover the application from many memory management related problems such as memory corruption and double free, certain non-deterministic software bugs such as race conditions, or user requests based problems that are triggered from some user requests or inputs.

ASSURE \cite{assure} identifies an application's rescue points, which are locations in the existing application code for handling failures, and inserts checkpoint and rollback code into the system with runtime binary injection. It first discovers candidate rescue points in an application offline by exploring the call-graphs of the application's executions when failure handling code is being triggered. An online test is then used to determine which candidate rescue point is most effective. The selected rescue point is inserted into the application in the production environment at runtime. When a fault occurs in the application, ASSURE restores execution to an appropriate rescue point and uses the applications own failure handling code to recovery from the fault.

Furthermore, work by others \cite{Chandra2000, Lowell2000} has shown that application-specific knowledge is necessary to recover from many kinds of application faults. Brown et al. build an e-mail store system \cite{Brown2003} that is capable of rolling back and replaying application-specific user operations. Each user operation is converted into verbs, which is an encapsulation of all the operations needed to re-execute or compensate a user operation, and is recorded into a timeline log on disk by an undo manager. During the troubleshooting process, the user can edit the timeline log to remove, replace, or add verbs. The undo manager then re-executes the appropriate part of the timeline. This mechanism reflects the Rewind, Repair, and Replay model \cite{Brown2002} that allows a user to repair a problem without losing work that was performed after the time at which the problem occurred.

\subsection{Versioning File Systems}
Versioning file systems keep a history of changes made to the file system to provide the ability to recover from file system related faults. Research in the field of versioning file systems can be classified into comprehensive versioning file systems, open-close versioning file systems, and snapshots versioning file systems,. 

Comprehensive versioning file systems \cite{soules:fast2003, wayback} create a new version for every write to the file system. This approach provides the finest granularity of versioning. Wayback \cite{wayback} is a user-level versioning file system built on the FUSE framework \cite{FUSE}. It logs the data to be overwritten or truncated into an undo log before any overwriting or truncating is performed. Our versioning file system uses an undo log similar to that of Wayback, but it is not implemented as a user-level file system. Rather it is implemented as a kernel module that intercepts file system calls made by user applications to create versions. The main drawback of a comprehensive versioning file system is the space overhead required since a great number of versions are created.

Open-close versioning file system \cite{ElephantFS} create a new version for all the writes occurs in a session of an open and a close of a file. The Elephant file system \cite{ElephantFS} is a versioning file system that provides users with a range of version retension policies. One interesting retention policy, called Keep Landmarks, retains only landmark versions in a file's history. A landmark version can either be specified by the user or automatically identified with heuristics, which is based on the assumption that versions generated within a short time period will be indistinguishable to the user if the versions were generated a long time ago.

Recently, Muniswamy-Reddy et al. proposed  a versioning file system \cite{Reddy2009} that leverages the causal relationships among files. It aims to gain the benefits of both open-close versioning file systems and comprehensive versioning file systems by creating a new version only when a new causal relationship is observed between a process and a file. This limits the number of versions to be created while preserving sufficient causality information for the recovery.

Snapshots versioning file systems \cite{Kistler1992, Peterson2005} create versions for the entire file system. Each version is a snapshot or an image of the file system. Due to the overhead of taking and saving file system snaphot, these systems usually create versions periodically to avoid disrupting normal user operations.

\subsection{Rabin fingerprints}
Rabin fingerprinting schema \cite{rabin_fingerprint} generates a fingerprint for a string by using randomly chosen irreducible polynomials. It guarantees the probability of a collision when two different strings have the same fingerprint can be made to be very small by choosing a sufficient number of bits to represent the fingerprint. It is widely used by string-matching and hashing algorithms since it is fast and is easy to implement.

Several systems have also used Rabin fingerprints to detect similarity
across files.  LBFS~\cite{lbfs} uses it to detect similar chunks among different files for the purposes of optimizing bandwidth usage in a
distributed files system. It divides the files it stores into chunks and indexes the chunks by the Rabin fingerprint of the chunks. It improves transmitting performance by avoiding transmitting a file chunk if any file on the recipient contains the same chunk. 
Forman et al.~\cite{HP:similarity} use it to
find similar documents to help manage large document repositories.
Rabin fingerprints were also used by Earlybird~\cite{earlybird} and
Autograph~\cite{autograph} to detect common substrings in network
traffic to detect Internet worms.
%\subsection{String Match}
%A Guided Tour to Approximate String Matching \cite{Navarro2001}
% conferences
% SOSP 2009 Done
% OSDI 2008 Done
% NSDI 2009 Done
% Usenix
%\end{comment}

%% file: conclusion.tex
\section{Conclusions}\label{section:conclusions}
Without knowledge of which files contain configuration state, trying
to revert a misconfigured application back to a working state is a
daunting task.  In addition to configuration state, applications
maintain a variety of task-dependent and time-dependent application
state strewn across files, all of which are difficult to distinguish
from configuration state.  SAIC solves this problem by using a
statistical approach that measures the similarity of a file over its
versions.  We find that with 3 simple filters that remove
non-persistent files, temporary files and user data files, and a
similarity metric that measures how chunks persist over the lifetime
of a file, SAIC is able to differentiate configuration files from
non-configuration files.  By identifying configuration files, it
reduces the number of files the user must examine for rollback by one
to two orders of magnitude.

SAIC can operate at various similarity thresholds, which are
implicitly determine by the amount of versioning storage space.
Depending on the threshold, it can identify between 60-100\% of
configuration files for most applications, while suffering a false
positive rate of 20-60\%.  With a threshold of 80\%, 90\% of
configuration files are correctly identified and 60\% of
non-configuration files are eliminated, removing an aggregate of
713129 file versions and 7GB of storage across two traces.  SAIC's
resource requirements are low due to two mechanisms.  A trigger point
function that determines an optimal measurement point by ensuring that
a file's versions have covered a sufficiently long period of time,
allows SAIC to take only one measurement per file.  Further, the
cost of similarity computation is reduced by using a sampling method
that performs the computation on fewer versions.  Both of these
enhancements have negligible effects on the accuracy of SAIC, so we
believe they should be used continuously during regular operation.

%% file: appendix.tex
\appendix
\section{Application Information}
%\section{Application Information}
\label{sec:appendix:apps}

{\bf Firefox} is a full featured web browser.  While Firefox spreads
its configuration over a large number of files, two of those
files\footnote{A reference for Firefox files can be found at {\tt
http://kb.mozillazine.org/Profile\_folder\_-\_Firefox}}, {\em
urlclassifier3.sqlite}, which contains a blacklist of phishing sites
and {\em pluginreg.dat}, which records associations of plugins with
MIME types account for over 95\% of the configuration file versions in
Trace 1 and over 99\% of the configuration file versions in Trace 2.
Strangely, we observed that Firefox frequently changes the order in
which the MIME associations appeared in the file, even though the
order of appearance has no effect on their semantic meaning -- an
example of the arbitrary and senseless behavior that all applications
exhibit to some degree.  While the meaning of the file contents  has
not changed, these updates cause the file contents to change and thus
results in a new version being created.  Some examples of Firefox
non-configuration files include  {\em sessionstore.js}, which records
task-dependent session information, which is used to reinitialize the
state of the browser after a crash, as well as files that store
task-dependent browsing state such as a history of form field values
and the user's cookie jar.

%A variety of state
%files exist as well, such as {\em sessionstore.js}, which records
%task-dependent session information, which is used to reinitiallize the
%state of the browser after a crash.  Firefox also stores a variety
%task-dependent history files, such as {\em formhistory.dat} and {\em
%cookiest.sqlite}.

{\bf GNOME} represents the GNOME desktop system,
 which includes session manager, window manager, power manager, 
 screen saver, GConf configuration system, as well as GNOME applications come with
 the default GNOME desktop installtion.
Instead of managing their own
configuration settings, many GNOME applications choose to use the GConf configuration
system to access, modify, and maintain their configurations. In both Trace 1 and Trace 2, 
all the GNOME configuration files were accessed and modified by GConf configuration system.

{\bf Flash} represents the Macromedia Flash plugin.  This plugin only
has one global configuration file and caches many website specific
settings.  These website specific settings are updated less frequently
than the global configuration file.

{\bf VMware} workstation is a popular hypervisor.  It has 3 config
files and the user in Trace 1 used it under two users for a total of 6
configuration files in the trace.  VMware creates a large number
of temporary lock files and log files.

{\bf JEdit} is a JAVA-based editor.  It has two configuration files
and several history files that store recently opened files and a
history of user actions.

{\bf Amarok} is an open-source music player that uses the KDE
framework.  It has two configuration files.  It stores user listening
habits and song rankings in an sqlite database file which we classify
as a non-configuration file because its contents are largely
task-dependent.

%% file: paper.bbl

\begin{thebibliography}{00}

%%% the Web address.

\ifx \showCODEN    \undefined \def \showCODEN     #1{\unskip}     \fi
\ifx \showDOI      \undefined \def \showDOI       #1{#1}\fi
\ifx \showISBNx    \undefined \def \showISBNx     #1{\unskip}     \fi
\ifx \showISBNxiii \undefined \def \showISBNxiii  #1{\unskip}     \fi
\ifx \showISSN     \undefined \def \showISSN      #1{\unskip}     \fi
\ifx \showLCCN     \undefined \def \showLCCN      #1{\unskip}     \fi
\ifx \shownote     \undefined \def \shownote      #1{#1}          \fi
\ifx \showarticletitle \undefined \def \showarticletitle #1{#1}   \fi
\ifx \showURL      \undefined \def \showURL       {\relax}        \fi
% invisible to TeX
\providecommand\bibfield[2]{#2}
\providecommand\bibinfo[2]{#2}
\providecommand\natexlab[1]{#1}
\providecommand\showeprint[2][]{arXiv:#2}

\bibitem[\protect\citeauthoryear{Aggarwal, Bhagwan, Das, Eswaran, Padmanabhan,
  and Voelker}{Aggarwal et~al\mbox{.}}{2009}]%
        {netprints}
\bibfield{author}{\bibinfo{person}{Bhavish Aggarwal}, \bibinfo{person}{Ranjita
  Bhagwan}, \bibinfo{person}{Tathagata Das}, \bibinfo{person}{Siddharth
  Eswaran}, \bibinfo{person}{Venkata~N. Padmanabhan}, {and}
  \bibinfo{person}{Geoffrey~M. Voelker}.} \bibinfo{year}{2009}\natexlab{}.
\newblock \showarticletitle{NetPrints: diagnosing home network
  misconfigurations using shared knowledge}. In \bibinfo{booktitle}{{\em
  NSDI'09: Proceedings of the 6th USENIX symposium on Networked systems design
  and implementation}}. \bibinfo{publisher}{USENIX Association},
  \bibinfo{address}{Berkeley, CA, USA}, \bibinfo{pages}{349--364}.
\newblock


\bibitem[\protect\citeauthoryear{Anderson, Goldsack, and Paterson}{Anderson
  et~al\mbox{.}}{2003}]%
        {Anderson2003}
\bibfield{author}{\bibinfo{person}{Paul Anderson}, \bibinfo{person}{Patrick
  Goldsack}, {and} \bibinfo{person}{Jim Paterson}.}
  \bibinfo{year}{2003}\natexlab{}.
\newblock \showarticletitle{SmartFrog Meets {LCFG}: Autonomous Reconfiguration
  with Central Policy Control}. In \bibinfo{booktitle}{{\em LISA '03:
  Proceedings of the 17th USENIX conference on System administration}}.
  \bibinfo{publisher}{USENIX Association}, \bibinfo{address}{Berkeley, CA,
  USA}, \bibinfo{pages}{213--222}.
\newblock


\bibitem[\protect\citeauthoryear{Anderson and Scobie}{Anderson and
  Scobie}{2002}]%
        {Anderson2002}
\bibfield{author}{\bibinfo{person}{P Anderson} {and} \bibinfo{person}{A
  Scobie}.} \bibinfo{year}{2002}\natexlab{}.
\newblock \showarticletitle{{LCFG}: The Next Generation}. In
  \bibinfo{booktitle}{{\em Proceedings of the UKUUG Winter Conference}}.
\newblock


\bibitem[\protect\citeauthoryear{Attariyan and Flinn}{Attariyan and
  Flinn}{2008}]%
        {Mona2008}
\bibfield{author}{\bibinfo{person}{Mona Attariyan} {and} \bibinfo{person}{Jason
  Flinn}.} \bibinfo{year}{2008}\natexlab{}.
\newblock \showarticletitle{Using causality to diagnose configuration bugs}. In
  \bibinfo{booktitle}{{\em ATC'08: USENIX 2008 Annual Technical Conference on
  Annual Technical Conference}}. \bibinfo{publisher}{USENIX Association},
  \bibinfo{address}{Berkeley, CA, USA}, \bibinfo{pages}{281--286}.
\newblock


\bibitem[\protect\citeauthoryear{Baker and Sullivan}{Baker and
  Sullivan}{1992}]%
        {Baker92}
\bibfield{author}{\bibinfo{person}{Mary Baker} {and} \bibinfo{person}{Mark
  Sullivan}.} \bibinfo{year}{1992}\natexlab{}.
\newblock \showarticletitle{The Recovery Box: Using Fast Recovery to Provide
  High Availability in the UNIX Environment}. In \bibinfo{booktitle}{{\em In
  Proceedings USENIX Summer Conference}}. \bibinfo{pages}{31--43}.
\newblock


\bibitem[\protect\citeauthoryear{Bressoud and Schneider}{Bressoud and
  Schneider}{1996}]%
        {Bressoud1996}
\bibfield{author}{\bibinfo{person}{Thomas~C. Bressoud} {and}
  \bibinfo{person}{Fred~B. Schneider}.} \bibinfo{year}{1996}\natexlab{}.
\newblock \showarticletitle{Hypervisor-based fault tolerance}.
\newblock \bibinfo{journal}{{\em ACM Trans. Comput. Syst.\/}}
  \bibinfo{volume}{14}, \bibinfo{number}{1} (\bibinfo{year}{1996}),
  \bibinfo{pages}{80--107}.
\newblock
\showISSN{0734-2071}
\showDOI{%
\url{https://doi.org/10.1145/225535.225538}}


\bibitem[\protect\citeauthoryear{Brown and Patterson}{Brown and
  Patterson}{2002}]%
        {Brown2002}
\bibfield{author}{\bibinfo{person}{Aaron~B. Brown} {and}
  \bibinfo{person}{David~A. Patterson}.} \bibinfo{year}{2002}\natexlab{}.
\newblock \showarticletitle{Rewind, repair, replay: three R's to
  dependability}. In \bibinfo{booktitle}{{\em EW10: Proceedings of the 10th
  workshop on ACM SIGOPS European workshop}}. \bibinfo{publisher}{ACM},
  \bibinfo{address}{New York, NY, USA}, \bibinfo{pages}{70--77}.
\newblock
\showDOI{%
\url{https://doi.org/10.1145/1133373.1133387}}


\bibitem[\protect\citeauthoryear{Brown and Patterson}{Brown and
  Patterson}{2003}]%
        {Brown2003}
\bibfield{author}{\bibinfo{person}{Aaron~B. Brown} {and}
  \bibinfo{person}{David~A. Patterson}.} \bibinfo{year}{2003}\natexlab{}.
\newblock \showarticletitle{Undo for operators: building an undoable e-mail
  store}. In \bibinfo{booktitle}{{\em ATEC '03: Proceedings of the annual
  conference on USENIX Annual Technical Conference}}.
  \bibinfo{publisher}{USENIX Association}, \bibinfo{address}{Berkeley, CA,
  USA}, \bibinfo{pages}{1--1}.
\newblock


\bibitem[\protect\citeauthoryear{Burgess}{Burgess}{1995}]%
        {Burgess1995}
\bibfield{author}{\bibinfo{person}{M. Burgess}.}
  \bibinfo{year}{1995}\natexlab{}.
\newblock \showarticletitle{Cfengine: a site configuration engine}.
\newblock \bibinfo{journal}{{\em USENIX Computing systems 8\/}}
  \bibinfo{volume}{3} (\bibinfo{year}{1995}).
\newblock


\bibitem[\protect\citeauthoryear{Candea, Kawamoto, Fujiki, Friedman, and
  Fox}{Candea et~al\mbox{.}}{2004}]%
        {Candea2004}
\bibfield{author}{\bibinfo{person}{George Candea}, \bibinfo{person}{Shinichi
  Kawamoto}, \bibinfo{person}{Yuichi Fujiki}, \bibinfo{person}{Greg Friedman},
  {and} \bibinfo{person}{Armando Fox}.} \bibinfo{year}{2004}\natexlab{}.
\newblock \showarticletitle{Microreboot --- A technique for cheap recovery}. In
  \bibinfo{booktitle}{{\em OSDI'04: Proceedings of the 6th conference on
  Symposium on Opearting Systems Design \& Implementation}}.
  \bibinfo{publisher}{USENIX Association}, \bibinfo{address}{Berkeley, CA,
  USA}, \bibinfo{pages}{3--3}.
\newblock


\bibitem[\protect\citeauthoryear{Chandra and Chen}{Chandra and Chen}{2000}]%
        {Chandra2000}
\bibfield{author}{\bibinfo{person}{S. Chandra} {and} \bibinfo{person}{P.M.
  Chen}.} \bibinfo{year}{2000}\natexlab{}.
\newblock \showarticletitle{Whither generic recovery from application faults? A
  fault study using open-source software}. In \bibinfo{booktitle}{{\em
  Dependable Systems and Networks, 2000. DSN 2000. Proceedings International
  Conference on}}. \bibinfo{pages}{97--106}.
\newblock
\showDOI{%
\url{https://doi.org/10.1109/ICDSN.2000.857521}}


\bibitem[\protect\citeauthoryear{Cormen, Leiserson, and Rivest}{Cormen
  et~al\mbox{.}}{1998}]%
        {clr}
\bibfield{author}{\bibinfo{person}{T. Cormen}, \bibinfo{person}{C. Leiserson},
  {and} \bibinfo{person}{R. Rivest}.} \bibinfo{year}{1998}\natexlab{}.
\newblock \bibinfo{booktitle}{{\em Introduction to Algorithms}}.
\newblock \bibinfo{publisher}{The MIT Press}, \bibinfo{pages}{312--314}.
\newblock


\bibitem[\protect\citeauthoryear{Cornell, Dinda, and Bustamante}{Cornell
  et~al\mbox{.}}{2004}]%
        {wayback}
\bibfield{author}{\bibinfo{person}{Brian Cornell}, \bibinfo{person}{Peter~A.
  Dinda}, {and} \bibinfo{person}{Fabi{\'a}n~E. Bustamante}.}
  \bibinfo{year}{2004}\natexlab{}.
\newblock \showarticletitle{Wayback: A User-level Versioning File System for
  {L}inux}. In \bibinfo{booktitle}{{\em Proceedings of the 2004 USENIX Annual
  Technical Conference, FREENIX Track}}. \bibinfo{pages}{19--28}.
\newblock


\bibitem[\protect\citeauthoryear{Crameri, Knezevic, Kostic, Bianchini, and
  Zwaenepoel}{Crameri et~al\mbox{.}}{2007}]%
        {mirage}
\bibfield{author}{\bibinfo{person}{Olivier Crameri}, \bibinfo{person}{Nikola
  Knezevic}, \bibinfo{person}{Dejan Kostic}, \bibinfo{person}{Ricardo
  Bianchini}, {and} \bibinfo{person}{Willy Zwaenepoel}.}
  \bibinfo{year}{2007}\natexlab{}.
\newblock \showarticletitle{Staged deployment in mirage, an integrated software
  upgrade testing and distribution system}. In \bibinfo{booktitle}{{\em SOSP
  '07: Proceedings of twenty-first ACM SIGOPS symposium on Operating systems
  principles}}. \bibinfo{publisher}{ACM}, \bibinfo{address}{New York, NY, USA},
  \bibinfo{pages}{221--236}.
\newblock
\showISBNx{978-1-59593-591-5}
\showDOI{%
\url{https://doi.org/10.1145/1294261.1294283}}


\bibitem[\protect\citeauthoryear{Dzsekijo}{Dzsekijo}{2003}]%
        {FUSE}
\bibfield{author}{\bibinfo{person}{Mszeredi Dzsekijo}.}
  \bibinfo{year}{2003}\natexlab{}.
\newblock \bibinfo{title}{{F}ilesystem in {U}serspace}.
\newblock \bibinfo{howpublished}{http://fuse.sourceforge.net}.
  (\bibinfo{year}{2003}).
\newblock


\bibitem[\protect\citeauthoryear{Elnozahy, Alvisi, Wang, and Johnson}{Elnozahy
  et~al\mbox{.}}{2002}]%
        {Elnozahy2002}
\bibfield{author}{\bibinfo{person}{E.~N.~(Mootaz) Elnozahy},
  \bibinfo{person}{Lorenzo Alvisi}, \bibinfo{person}{Yi-Min Wang}, {and}
  \bibinfo{person}{David~B. Johnson}.} \bibinfo{year}{2002}\natexlab{}.
\newblock \showarticletitle{A survey of rollback-recovery protocols in
  message-passing systems}.
\newblock \bibinfo{journal}{{\em ACM Comput. Surv.\/}} \bibinfo{volume}{34},
  \bibinfo{number}{3} (\bibinfo{year}{2002}), \bibinfo{pages}{375--408}.
\newblock
\showISSN{0360-0300}
\showDOI{%
\url{https://doi.org/10.1145/568522.568525}}


\bibitem[\protect\citeauthoryear{Forman, Eshghi, and Chiocchetti}{Forman
  et~al\mbox{.}}{2005}]%
        {HP:similarity}
\bibfield{author}{\bibinfo{person}{George Forman}, \bibinfo{person}{Kave
  Eshghi}, {and} \bibinfo{person}{Stephane Chiocchetti}.}
  \bibinfo{year}{2005}\natexlab{}.
\newblock \showarticletitle{Finding similar files in large document
  repositories}. In \bibinfo{booktitle}{{\em Proceedings of the 11th {ACM}
  {SIGKDD} International Conference on Knowledge Discovery in Data Mining
  ({KDD})}}. \bibinfo{pages}{394--400}.
\newblock


\bibitem[\protect\citeauthoryear{Goel, Farhadi, Po, and Feng}{Goel
  et~al\mbox{.}}{2008}]%
        {Goel2008}
\bibfield{author}{\bibinfo{person}{Ashvin Goel}, \bibinfo{person}{Kamran
  Farhadi}, \bibinfo{person}{Kenneth Po}, {and} \bibinfo{person}{Wu-chang
  Feng}.} \bibinfo{year}{2008}\natexlab{}.
\newblock \showarticletitle{Reconstructing system state for intrusion
  analysis}.
\newblock \bibinfo{journal}{{\em SIGOPS Oper. Syst. Rev.\/}}
  \bibinfo{volume}{42}, \bibinfo{number}{3} (\bibinfo{year}{2008}),
  \bibinfo{pages}{21--28}.
\newblock
\showISSN{0163-5980}
\showDOI{%
\url{https://doi.org/10.1145/1368506.1368511}}


\bibitem[\protect\citeauthoryear{Gray}{Gray}{1986}]%
        {Gray1986}
\bibfield{author}{\bibinfo{person}{Jim Gray}.} \bibinfo{year}{1986}\natexlab{}.
\newblock \showarticletitle{Why Do Computers Stop and What Can Be Done About
  It?}. In \bibinfo{booktitle}{{\em Symposium on Reliability in Distributed
  Software and Database Systems}}. \bibinfo{pages}{3--12}.
\newblock
\showURL{%
\url{http://citeseerx.ist.psu.edu/viewdoc/summary?doi=10.1.1.59.6561}}


\bibitem[\protect\citeauthoryear{Gray}{Gray}{1990}]%
        {Gray1990}
\bibfield{author}{\bibinfo{person}{Jim Gray}.} \bibinfo{year}{1990}\natexlab{}.
\newblock \showarticletitle{A Census of Tandem System Availability Between 1985
  and 1990}, Vol.~\bibinfo{volume}{39}. IEEE, \bibinfo{pages}{409--418}.
\newblock
\showURL{%
\url{http://ieeexplore-beta.ieee.org//iel1/24/2133/00058719.pdf}}


\bibitem[\protect\citeauthoryear{Inc.}{Inc.}{2009}]%
        {timemachine}
\bibfield{author}{\bibinfo{person}{Apple Inc.}}
  \bibinfo{year}{2009}\natexlab{}.
\newblock \bibinfo{title}{{W}hat is {Mac} {OS X} - {T}Ime {M}achine}.
\newblock
  \bibinfo{howpublished}{http://www.apple.com/macosx/what-is-macosx/time-machine.html}.
    (\bibinfo{year}{2009}).
\newblock
\newblock
\shownote{Last accessed: 9/1/2009.}


\bibitem[\protect\citeauthoryear{Jiang, Hu, Pasupathy, Kanevsky, Li, and
  Zhou}{Jiang et~al\mbox{.}}{2009}]%
        {Jiang2009}
\bibfield{author}{\bibinfo{person}{Weihang Jiang}, \bibinfo{person}{Chongfeng
  Hu}, \bibinfo{person}{Shankar Pasupathy}, \bibinfo{person}{Arkady Kanevsky},
  \bibinfo{person}{Zhenmin Li}, {and} \bibinfo{person}{Yuanyuan Zhou}.}
  \bibinfo{year}{2009}\natexlab{}.
\newblock \showarticletitle{Understanding customer problem troubleshooting from
  storage system logs}. In \bibinfo{booktitle}{{\em FAST '09: Proccedings of
  the 7th conference on File and storage technologies}}.
  \bibinfo{publisher}{USENIX Association}, \bibinfo{address}{Berkeley, CA,
  USA}, \bibinfo{pages}{43--56}.
\newblock


\bibitem[\protect\citeauthoryear{Kim and Karp}{Kim and Karp}{2004}]%
        {autograph}
\bibfield{author}{\bibinfo{person}{Hyang-Ah Kim} {and} \bibinfo{person}{Brad
  Karp}.} \bibinfo{year}{2004}\natexlab{}.
\newblock \showarticletitle{Autograph: Toward automated, distributed worm
  signature detection}. In \bibinfo{booktitle}{{\em Proc. of the 13th {USENIX}
  Security Symposium}}. \bibinfo{pages}{271--286}.
\newblock


\bibitem[\protect\citeauthoryear{King, Dunlap, and Chen}{King
  et~al\mbox{.}}{2005}]%
        {King2005}
\bibfield{author}{\bibinfo{person}{Samuel~T. King}, \bibinfo{person}{George~W.
  Dunlap}, {and} \bibinfo{person}{Peter~M. Chen}.}
  \bibinfo{year}{2005}\natexlab{}.
\newblock \showarticletitle{Debugging operating systems with time-traveling
  virtual machines}. In \bibinfo{booktitle}{{\em ATEC '05: Proceedings of the
  annual conference on USENIX Annual Technical Conference}}.
  \bibinfo{publisher}{USENIX Association}, \bibinfo{address}{Berkeley, CA,
  USA}, \bibinfo{pages}{1--1}.
\newblock


\bibitem[\protect\citeauthoryear{Kistler and Satyanarayanan}{Kistler and
  Satyanarayanan}{1992}]%
        {Kistler1992}
\bibfield{author}{\bibinfo{person}{James~J. Kistler} {and} \bibinfo{person}{M.
  Satyanarayanan}.} \bibinfo{year}{1992}\natexlab{}.
\newblock \showarticletitle{Disconnected operation in the Coda File System}.
\newblock \bibinfo{journal}{{\em ACM Trans. Comput. Syst.\/}}
  \bibinfo{volume}{10}, \bibinfo{number}{1} (\bibinfo{year}{1992}),
  \bibinfo{pages}{3--25}.
\newblock
\showISSN{0734-2071}
\showDOI{%
\url{https://doi.org/10.1145/146941.146942}}


\bibitem[\protect\citeauthoryear{Kolettis and Fulton}{Kolettis and
  Fulton}{1995}]%
        {Kolettis1995}
\bibfield{author}{\bibinfo{person}{Nick Kolettis} {and}
  \bibinfo{person}{N.~Dudley Fulton}.} \bibinfo{year}{1995}\natexlab{}.
\newblock \showarticletitle{Software Rejuvenation: Analysis, Module and
  Applications}. In \bibinfo{booktitle}{{\em FTCS '95: Proceedings of the
  Twenty-Fifth International Symposium on Fault-Tolerant Computing}}.
  \bibinfo{publisher}{IEEE Computer Society}, \bibinfo{address}{Washington, DC,
  USA}, \bibinfo{pages}{381}.
\newblock


\bibitem[\protect\citeauthoryear{Laadan and Nieh}{Laadan and Nieh}{2007}]%
        {Laadan2007}
\bibfield{author}{\bibinfo{person}{Oren Laadan} {and} \bibinfo{person}{Jason
  Nieh}.} \bibinfo{year}{2007}\natexlab{}.
\newblock \showarticletitle{Transparent checkpoint-restart of multiple
  processes on commodity operating systems}. In \bibinfo{booktitle}{{\em
  ATC'07: 2007 USENIX Annual Technical Conference on Proceedings of the USENIX
  Annual Technical Conference}}. \bibinfo{publisher}{USENIX Association},
  \bibinfo{address}{Berkeley, CA, USA}, \bibinfo{pages}{1--14}.
\newblock
\showISBNx{999-8888-77-6}


\bibitem[\protect\citeauthoryear{Lenharth, Adve, and King}{Lenharth
  et~al\mbox{.}}{2009}]%
        {Lenharth2009}
\bibfield{author}{\bibinfo{person}{Andrew Lenharth}, \bibinfo{person}{Vikram~S.
  Adve}, {and} \bibinfo{person}{Samuel~T. King}.}
  \bibinfo{year}{2009}\natexlab{}.
\newblock \showarticletitle{Recovery domains: an organizing principle for
  recoverable operating systems}. In \bibinfo{booktitle}{{\em ASPLOS '09:
  Proceeding of the 14th international conference on Architectural support for
  programming languages and operating systems}}. \bibinfo{publisher}{ACM},
  \bibinfo{address}{New York, NY, USA}, \bibinfo{pages}{49--60}.
\newblock
\showISBNx{978-1-60558-406-5}
\showDOI{%
\url{https://doi.org/10.1145/1508244.1508251}}


\bibitem[\protect\citeauthoryear{Lowell, Chandra, and Chen}{Lowell
  et~al\mbox{.}}{2000}]%
        {Lowell2000}
\bibfield{author}{\bibinfo{person}{David~E. Lowell},
  \bibinfo{person}{Subhachandra Chandra}, {and} \bibinfo{person}{Peter~M.
  Chen}.} \bibinfo{year}{2000}\natexlab{}.
\newblock \showarticletitle{Exploring failure transparency and the limits of
  generic recovery}. In \bibinfo{booktitle}{{\em OSDI'00: Proceedings of the
  4th conference on Symposium on Operating System Design \& Implementation}}.
  \bibinfo{publisher}{USENIX Association}, \bibinfo{address}{Berkeley, CA,
  USA}, \bibinfo{pages}{20--20}.
\newblock


\bibitem[\protect\citeauthoryear{Marcus and Stern}{Marcus and Stern}{2000}]%
        {Marcus2000}
\bibfield{author}{\bibinfo{person}{E. Marcus} {and} \bibinfo{person}{H.
  Stern}.} \bibinfo{year}{2000}\natexlab{}.
\newblock \bibinfo{booktitle}{{\em Blueprints for High Availability}}.
\newblock \bibinfo{publisher}{John Willey \& Sons}.
\newblock


\bibitem[\protect\citeauthoryear{Mickens, Szummer, and Narayanan}{Mickens
  et~al\mbox{.}}{2007}]%
        {snitch}
\bibfield{author}{\bibinfo{person}{James Mickens}, \bibinfo{person}{Martin
  Szummer}, {and} \bibinfo{person}{Dushyanth Narayanan}.}
  \bibinfo{year}{2007}\natexlab{}.
\newblock \showarticletitle{Snitch: interactive decision trees for
  troubleshooting misconfigurations}. In \bibinfo{booktitle}{{\em SYSML'07:
  Proceedings of the 2nd USENIX workshop on Tackling computer systems problems
  with machine learning techniques}}. \bibinfo{publisher}{USENIX Association},
  \bibinfo{address}{Berkeley, CA, USA}, \bibinfo{pages}{1--6}.
\newblock


\bibitem[\protect\citeauthoryear{{M}icrosoft {C}orporation.}{{M}icrosoft
  {C}orporation.}{2009}]%
        {shadowcopy}
\bibfield{author}{\bibinfo{person}{{M}icrosoft {C}orporation.}}
  \bibinfo{year}{2009}\natexlab{}.
\newblock \bibinfo{title}{Explore the features: Shadow Copy}.
\newblock
  \bibinfo{howpublished}{http://www.microsoft.com/windows/windows-vista/features/shadow-copy.aspx}.
    (\bibinfo{year}{2009}).
\newblock
\newblock
\shownote{Last accessed: 9/1/2009.}


\bibitem[\protect\citeauthoryear{Muniswamy-Reddy and Holland}{Muniswamy-Reddy
  and Holland}{2009}]%
        {Reddy2009}
\bibfield{author}{\bibinfo{person}{Kiran-Kumar Muniswamy-Reddy} {and}
  \bibinfo{person}{David~A. Holland}.} \bibinfo{year}{2009}\natexlab{}.
\newblock \showarticletitle{Causality-based versioning}. In
  \bibinfo{booktitle}{{\em FAST '09: Proccedings of the 7th conference on File
  and storage technologies}}. \bibinfo{publisher}{USENIX Association},
  \bibinfo{address}{Berkeley, CA, USA}, \bibinfo{pages}{15--28}.
\newblock


\bibitem[\protect\citeauthoryear{Muniswamy-Reddy, Wright, Himmer, and
  Zadok}{Muniswamy-Reddy et~al\mbox{.}}{2004}]%
        {versionfs}
\bibfield{author}{\bibinfo{person}{Kiran-Kumar Muniswamy-Reddy},
  \bibinfo{person}{C.~P. Wright}, \bibinfo{person}{A. Himmer}, {and}
  \bibinfo{person}{E. Zadok}.} \bibinfo{year}{2004}\natexlab{}.
\newblock \showarticletitle{{A Versatile and User-Oriented Versioning File
  System}}. \bibinfo{pages}{115--128}.
\newblock


\bibitem[\protect\citeauthoryear{Muthitacharoen, Chen, and
  Mazieres}{Muthitacharoen et~al\mbox{.}}{2001}]%
        {lbfs}
\bibfield{author}{\bibinfo{person}{Athicha Muthitacharoen},
  \bibinfo{person}{Benjie Chen}, {and} \bibinfo{person}{David Mazieres}.}
  \bibinfo{year}{2001}\natexlab{}.
\newblock \showarticletitle{A Low-Bandwidth Network File System}. In
  \bibinfo{booktitle}{{\em Proc. of the 18th {ACM} {S}ymposium on {O}perating
  {S}ystems {P}rinciples}}. \bibinfo{pages}{174--187}.
\newblock


\bibitem[\protect\citeauthoryear{Nagaraja, Oliveira, Bianchini, Martin, and
  Nguyen}{Nagaraja et~al\mbox{.}}{2004}]%
        {Nagaraja2004}
\bibfield{author}{\bibinfo{person}{Kiran Nagaraja}, \bibinfo{person}{F\'{a}bio
  Oliveira}, \bibinfo{person}{Ricardo Bianchini}, \bibinfo{person}{Richard~P.
  Martin}, {and} \bibinfo{person}{Thu~D. Nguyen}.}
  \bibinfo{year}{2004}\natexlab{}.
\newblock \showarticletitle{Understanding and dealing with operator mistakes in
  internet services}. In \bibinfo{booktitle}{{\em OSDI'04: Proceedings of the
  6th conference on Symposium on Opearting Systems Design \& Implementation}}.
  \bibinfo{publisher}{USENIX Association}, \bibinfo{address}{Berkeley, CA,
  USA}, \bibinfo{pages}{5--5}.
\newblock


\bibitem[\protect\citeauthoryear{Oliveira, Nagaraja, Bachwani, Bianchini,
  Martin, and Nguyen}{Oliveira et~al\mbox{.}}{2006}]%
        {Oliveira2006}
\bibfield{author}{\bibinfo{person}{F\'{a}bio Oliveira}, \bibinfo{person}{Kiran
  Nagaraja}, \bibinfo{person}{Rekha Bachwani}, \bibinfo{person}{Ricardo
  Bianchini}, \bibinfo{person}{Richard~P. Martin}, {and}
  \bibinfo{person}{Thu~D. Nguyen}.} \bibinfo{year}{2006}\natexlab{}.
\newblock \showarticletitle{Understanding and validating database system
  administration}. In \bibinfo{booktitle}{{\em ATEC '06: Proceedings of the
  annual conference on USENIX '06 Annual Technical Conference}}.
  \bibinfo{publisher}{USENIX Association}, \bibinfo{address}{Berkeley, CA,
  USA}, \bibinfo{pages}{19--19}.
\newblock


\bibitem[\protect\citeauthoryear{Oppenheimer, Ganapathi, and
  Patterson}{Oppenheimer et~al\mbox{.}}{2003}]%
        {Oppenheimer2003}
\bibfield{author}{\bibinfo{person}{David Oppenheimer}, \bibinfo{person}{Archana
  Ganapathi}, {and} \bibinfo{person}{David~A. Patterson}.}
  \bibinfo{year}{2003}\natexlab{}.
\newblock \showarticletitle{Why do internet services fail, and what can be done
  about it?}. In \bibinfo{booktitle}{{\em USITS'03: Proceedings of the 4th
  conference on USENIX Symposium on Internet Technologies and Systems}}.
  \bibinfo{publisher}{USENIX Association}, \bibinfo{address}{Berkeley, CA,
  USA}, \bibinfo{pages}{1--1}.
\newblock


\bibitem[\protect\citeauthoryear{Osman, Subhraveti, Su, and Nieh}{Osman
  et~al\mbox{.}}{2002}]%
        {zap}
\bibfield{author}{\bibinfo{person}{Steven Osman}, \bibinfo{person}{Dinesh
  Subhraveti}, \bibinfo{person}{Gong Su}, {and} \bibinfo{person}{Jason Nieh}.}
  \bibinfo{year}{2002}\natexlab{}.
\newblock \showarticletitle{The design and implementation of Zap: a system for
  migrating computing environments}. In \bibinfo{booktitle}{{\em OSDI '02:
  Proceedings of the 5th symposium on Operating systems design and
  implementation}}. \bibinfo{publisher}{ACM}, \bibinfo{address}{New York, NY,
  USA}, \bibinfo{pages}{361--376}.
\newblock
\showDOI{%
\url{https://doi.org/10.1145/1060289.1060323}}


\bibitem[\protect\citeauthoryear{Peterson and Burns}{Peterson and
  Burns}{2005}]%
        {Peterson2005}
\bibfield{author}{\bibinfo{person}{Zachary Peterson} {and}
  \bibinfo{person}{Randal Burns}.} \bibinfo{year}{2005}\natexlab{}.
\newblock \showarticletitle{Ext3cow: a time-shifting file system for regulatory
  compliance}.
\newblock \bibinfo{journal}{{\em Trans. Storage\/}} \bibinfo{volume}{1},
  \bibinfo{number}{2} (\bibinfo{year}{2005}), \bibinfo{pages}{190--212}.
\newblock
\showISSN{1553-3077}
\showDOI{%
\url{https://doi.org/10.1145/1063786.1063789}}


\bibitem[\protect\citeauthoryear{Qin, Tucek, Sundaresan, and Zhou}{Qin
  et~al\mbox{.}}{2005}]%
        {Feng2005}
\bibfield{author}{\bibinfo{person}{Feng Qin}, \bibinfo{person}{Joseph Tucek},
  \bibinfo{person}{Jagadeesan Sundaresan}, {and} \bibinfo{person}{Yuanyuan
  Zhou}.} \bibinfo{year}{2005}\natexlab{}.
\newblock \showarticletitle{Rx: treating bugs as allergies---a safe method to
  survive software failures}. In \bibinfo{booktitle}{{\em SOSP '05: Proceedings
  of the twentieth ACM symposium on Operating systems principles}}.
  \bibinfo{publisher}{ACM}, \bibinfo{address}{New York, NY, USA},
  \bibinfo{pages}{235--248}.
\newblock
\showISBNx{1-59593-079-5}
\showDOI{%
\url{https://doi.org/10.1145/1095810.1095833}}


\bibitem[\protect\citeauthoryear{Rabin}{Rabin}{1981}]%
        {rabin_fingerprint}
\bibfield{author}{\bibinfo{person}{Michael~O. Rabin}.}
  \bibinfo{year}{1981}\natexlab{}.
\newblock \bibinfo{booktitle}{{\em Fingerprinting by Random Polynomials}}.
\newblock \bibinfo{type}{{T}echnical {R}eport} TR-15-81.
  \bibinfo{institution}{Center for Research in Computer Technology, Harvard
  University}.
\newblock


\bibitem[\protect\citeauthoryear{Santry, Feeley, Hutchinson, Veitch, Carton,
  and Ofir}{Santry et~al\mbox{.}}{1999}]%
        {ElephantFS}
\bibfield{author}{\bibinfo{person}{Douglas~S. Santry},
  \bibinfo{person}{Michael~J. Feeley}, \bibinfo{person}{Norman~C. Hutchinson},
  \bibinfo{person}{Alistair~C. Veitch}, \bibinfo{person}{Ross~W. Carton}, {and}
  \bibinfo{person}{Jacob Ofir}.} \bibinfo{year}{1999}\natexlab{}.
\newblock \showarticletitle{Deciding when to forget in the {E}lephant file
  system}. In \bibinfo{booktitle}{{\em Proc. of the 17th {ACM} {S}ymposium on
  {O}perating {S}ystems {P}rinciples}}. \bibinfo{pages}{110--123}.
\newblock


\bibitem[\protect\citeauthoryear{Sidiroglou, Laadan, Perez, Viennot, Nieh, and
  Keromytis}{Sidiroglou et~al\mbox{.}}{2009}]%
        {assure}
\bibfield{author}{\bibinfo{person}{Stelios Sidiroglou}, \bibinfo{person}{Oren
  Laadan}, \bibinfo{person}{Carlos Perez}, \bibinfo{person}{Nicolas Viennot},
  \bibinfo{person}{Jason Nieh}, {and} \bibinfo{person}{Angelos~D. Keromytis}.}
  \bibinfo{year}{2009}\natexlab{}.
\newblock \showarticletitle{ASSURE: automatic software self-healing using
  rescue points}. In \bibinfo{booktitle}{{\em ASPLOS '09: Proceeding of the
  14th international conference on Architectural support for programming
  languages and operating systems}}. \bibinfo{publisher}{ACM},
  \bibinfo{address}{New York, NY, USA}, \bibinfo{pages}{37--48}.
\newblock
\showISBNx{978-1-60558-406-5}
\showDOI{%
\url{https://doi.org/10.1145/1508244.1508250}}


\bibitem[\protect\citeauthoryear{Singh, Estan, Varghese, and Savage}{Singh
  et~al\mbox{.}}{2004}]%
        {earlybird}
\bibfield{author}{\bibinfo{person}{Sumeet Singh}, \bibinfo{person}{Cristian
  Estan}, \bibinfo{person}{George Varghese}, {and} \bibinfo{person}{Stefan
  Savage}.} \bibinfo{year}{2004}\natexlab{}.
\newblock \showarticletitle{Automated Worm Fingerprinting}. In
  \bibinfo{booktitle}{{\em Proceedings of the 6th Symposium on Operating
  Systems Design and Implementation}}. \bibinfo{pages}{45--60}.
\newblock


\bibitem[\protect\citeauthoryear{Soules, Goodson, Strunk, and Ganger}{Soules
  et~al\mbox{.}}{2003}]%
        {soules:fast2003}
\bibfield{author}{\bibinfo{person}{Craig A.~N. Soules},
  \bibinfo{person}{Garth~R. Goodson}, \bibinfo{person}{John~D. Strunk}, {and}
  \bibinfo{person}{Gregory~R. Ganger}.} \bibinfo{year}{2003}\natexlab{}.
\newblock \showarticletitle{Metadata Efficiency in Versioning File Systems}. In
  \bibinfo{booktitle}{{\em FAST '03: Proceedings of the 2nd USENIX Conference
  on File and Storage Technologies}}. \bibinfo{publisher}{USENIX Association},
  \bibinfo{address}{Berkeley, CA, USA}, \bibinfo{pages}{43--58}.
\newblock


\bibitem[\protect\citeauthoryear{Su, Attariyan, and Flinn}{Su
  et~al\mbox{.}}{2007}]%
        {autobash}
\bibfield{author}{\bibinfo{person}{Ya-Yunn Su}, \bibinfo{person}{Mona
  Attariyan}, {and} \bibinfo{person}{Jason Flinn}.}
  \bibinfo{year}{2007}\natexlab{}.
\newblock \showarticletitle{AutoBash: improving configuration management with
  operating system causality analysis}. In \bibinfo{booktitle}{{\em Proceedings
  of the 21st {ACM} Symposium on Operating Systems Principles}}.
  \bibinfo{pages}{237--250}.
\newblock


\bibitem[\protect\citeauthoryear{Su and Flinn}{Su and Flinn}{2009}]%
        {Su2009}
\bibfield{author}{\bibinfo{person}{Ya-Yunn Su} {and} \bibinfo{person}{Jason
  Flinn}.} \bibinfo{year}{2009}\natexlab{}.
\newblock \showarticletitle{Automatically Generating Predicates and Solutions
  for Configuration Troubleshooting}. In \bibinfo{booktitle}{{\em ATC'09:
  USENIX 2009 Annual Technical Conference on Annual Technical Conference}}.
  \bibinfo{publisher}{USENIX Association}, \bibinfo{address}{Berkeley, CA,
  USA}.
\newblock


\bibitem[\protect\citeauthoryear{Sullivan and Chillarege}{Sullivan and
  Chillarege}{1991}]%
        {Sullivan1991}
\bibfield{author}{\bibinfo{person}{M. Sullivan} {and} \bibinfo{person}{R.
  Chillarege}.} \bibinfo{year}{1991}\natexlab{}.
\newblock \showarticletitle{Software defects and their impact on system
  availability - a study of field failures in operating systems}.
\newblock \bibinfo{journal}{{\em 21st Int. Symp. on Fault-Tolerant Computing
  (FTCS-21)\/}} (\bibinfo{year}{1991}), \bibinfo{pages}{2--9}.
\newblock
\showURL{%
\url{citeseer.ist.psu.edu/sullivan91software.html}}


\bibitem[\protect\citeauthoryear{Vaidyanathan and Trivedi}{Vaidyanathan and
  Trivedi}{2005}]%
        {Vaidyanathan2005}
\bibfield{author}{\bibinfo{person}{Kalyanaraman Vaidyanathan} {and}
  \bibinfo{person}{Kishor~S. Trivedi}.} \bibinfo{year}{2005}\natexlab{}.
\newblock \showarticletitle{A Comprehensive Model for Software Rejuvenation}.
\newblock \bibinfo{journal}{{\em IEEE Trans. Dependable Secur. Comput.\/}}
  \bibinfo{volume}{2}, \bibinfo{number}{2} (\bibinfo{year}{2005}),
  \bibinfo{pages}{124--137}.
\newblock
\showISSN{1545-5971}
\showDOI{%
\url{https://doi.org/10.1109/TDSC.2005.15}}
\newblock
\shownote{Member-Vaidyanathan, Kalyanaraman and Fellow-Trivedi, Kishor S.}


\bibitem[\protect\citeauthoryear{Wang, Platt, Chen, Zhang, and Wang}{Wang
  et~al\mbox{.}}{2004}]%
        {peerpressure}
\bibfield{author}{\bibinfo{person}{Helen~J. Wang}, \bibinfo{person}{John~C.
  Platt}, \bibinfo{person}{Yu Chen}, \bibinfo{person}{Ruyun Zhang}, {and}
  \bibinfo{person}{Yi-Min Wang}.} \bibinfo{year}{2004}\natexlab{}.
\newblock \showarticletitle{Automatic Misconfiguration Troubleshooting with
  {P}eer{P}ressure}. In \bibinfo{booktitle}{{\em Proceedings of the 6th
  Symposium on Operating Systems Design and Implementation}}.
  \bibinfo{pages}{245--258}.
\newblock


\bibitem[\protect\citeauthoryear{Wang, Verbowski, Dunagan, Chen, Wang, Yuan,
  and Zhang}{Wang et~al\mbox{.}}{2003}]%
        {strider}
\bibfield{author}{\bibinfo{person}{Yi-Min Wang}, \bibinfo{person}{Chad
  Verbowski}, \bibinfo{person}{John Dunagan}, \bibinfo{person}{Yu Chen},
  \bibinfo{person}{Helen~J. Wang}, \bibinfo{person}{Chun Yuan}, {and}
  \bibinfo{person}{Zheng Zhang}.} \bibinfo{year}{2003}\natexlab{}.
\newblock \showarticletitle{STRIDER: A Black-box, State-based Approach to
  Change and Configuration Management and Support}. In \bibinfo{booktitle}{{\em
  LISA '03: Proceedings of the 17th USENIX conference on System
  administration}}. \bibinfo{publisher}{USENIX Association},
  \bibinfo{address}{Berkeley, CA, USA}, \bibinfo{pages}{159--172}.
\newblock


\bibitem[\protect\citeauthoryear{Whitaker, Cox, and Gribble}{Whitaker
  et~al\mbox{.}}{2004}]%
        {chronus}
\bibfield{author}{\bibinfo{person}{Andrew Whitaker},
  \bibinfo{person}{Richard~S. Cox}, {and} \bibinfo{person}{Steven~D. Gribble}.}
  \bibinfo{year}{2004}\natexlab{}.
\newblock \showarticletitle{Configuration Debugging as Search: Finding the
  Needle in the Haystack.}. In \bibinfo{booktitle}{{\em Proceedings of the 6th
  Symposium on Operating Systems Design and Implementation}}.
  \bibinfo{pages}{77--90}.
\newblock


\bibitem[\protect\citeauthoryear{Yuan, Lao, Wen, Li, Zhang, Wang, and Ma}{Yuan
  et~al\mbox{.}}{2006}]%
        {Yuan2006}
\bibfield{author}{\bibinfo{person}{Chun Yuan}, \bibinfo{person}{Ni Lao},
  \bibinfo{person}{Ji-Rong Wen}, \bibinfo{person}{Jiwei Li},
  \bibinfo{person}{Zheng Zhang}, \bibinfo{person}{Yi-Min Wang}, {and}
  \bibinfo{person}{Wei-Ying Ma}.} \bibinfo{year}{2006}\natexlab{}.
\newblock \showarticletitle{Automated Known Problem Diagnosis with Event
  Traces}.
\newblock \bibinfo{journal}{{\em SIGOPS Oper. Syst. Rev.\/}}
  \bibinfo{volume}{40}, \bibinfo{number}{4}, \bibinfo{pages}{375--388}.
\newblock
\showISSN{0163-5980}
\showDOI{%
\url{https://doi.org/10.1145/1218063.1217972}}


\bibitem[\protect\citeauthoryear{Zheng, Bianchini, and Nguyen}{Zheng
  et~al\mbox{.}}{2007}]%
        {Zheng2007}
\bibfield{author}{\bibinfo{person}{Wei Zheng}, \bibinfo{person}{Ricardo
  Bianchini}, {and} \bibinfo{person}{Thu~D. Nguyen}.}
  \bibinfo{year}{2007}\natexlab{}.
\newblock \showarticletitle{Automatic configuration of internet services}. In
  \bibinfo{booktitle}{{\em EuroSys '07: Proceedings of the 2nd ACM
  SIGOPS/EuroSys European Conference on Computer Systems 2007}}.
  \bibinfo{publisher}{ACM}, \bibinfo{address}{New York, NY, USA},
  \bibinfo{pages}{219--229}.
\newblock
\showISBNx{978-1-59593-636-3}
\showDOI{%
\url{https://doi.org/10.1145/1272996.1273020}}


\end{thebibliography}
